\documentclass[article]{aa}  
\pdfoutput=1
\bibpunct{(}{)}{;}{a}{}{,} 
\usepackage[utf8]{inputenc}
\usepackage[switch]{lineno}

\usepackage[T1]{fontenc}
\usepackage{textcomp}
\usepackage{graphicx}
\usepackage{multirow}
\usepackage{pifont}
\usepackage{makecell} 
\setcellgapes{5pt}
\usepackage{txfonts}
\usepackage{rotating}
\usepackage[]{xcolor}
\usepackage{comment}
\usepackage{amsmath,bm}
\usepackage{booktabs}
\usepackage{nicematrix}
\usepackage{hyperref}

\newcommand{\hii}{\mbox{H\,{\sc{ii}}\, }}
\usepackage{graphicx}
\usepackage{txfonts}

\usepackage{tikz}
\usetikzlibrary{calc,positioning}
\usepackage{ulem}
\usepackage{subfigure}
\usepackage{ifthen}
\usepackage{placeins}
\usepackage{lipsum}
\usepackage{float}
\usepackage{xcolor}
\usepackage{xspace}

\hypersetup{colorlinks=true,linkcolor=blue,citecolor=blue,filecolor=blue,urlcolor=blue}
\graphicspath{{images/}}

\newcommand{\machav}{\mathcal{M}_\mathrm{A}}

\newcommand{\disperse}{\texttt{DisPerSE}\xspace}

\newcommand{\simBB}{{\tt High\_B}\xspace}
\newcommand{\simA}{{\tt Low\_B}\xspace}
\newcommand{\SFE}{\mathrm{SFE}}

\newcommand{\vpar}{V_\parallel}
\newcommand{\vperp}{V_\perp}

\newcommand{\kms}{\mathrm{km\,s^{-1}}}

\newcommand{\pc}{\mathrm{pc}}
\newcommand{\MSun}{\mathrm{M}_\odot}

\newcommand{\Myr}{\mathrm{Myr}}

\newcommand{\ndens}{\mathrm{cm^{-3}}} 
\newcommand{\dens}{\mathrm{g\,cm^{-3}}} 

\newcommand{\ncoldens}{\mathrm{cm^{-2}}}

\newboolean{ActivateComments}
\setboolean{ActivateComments}{false}

\newboolean{ActivatePaolosComments}
\setboolean{ActivatePaolosComments}{true}

\makeatletter
\newcommand*\bigcdot{\mathpalette\bigcdot@{.5}}
\newcommand*\bigcdot@[2]{\mathbin{\vcenter{\hbox{\scalebox{#2}{$\m@th#1\bullet$}}}}}
\makeatother

\DeclareRobustCommand{\Pcomment}[2]{%
  \ifthenelse{\boolean{ActivatePaolosComments}}{%
    \ifthenelse{\equal{#1}{}}{\textbf{#2}}{\textcolor{gray}{\sout{#1}} \textbf{#2}}}{#2}}

\newcommand{\Acomment}[1]{%
  \ifbool{ActivateComments}{\textcolor{olive}{#1}}{}%
}
\newcommand{\Dcomment}[1]{%
  \ifbool{ActivateComments}{\textcolor{cyan}{#1}}{}%
}

\definecolor{lime}{HTML}{A6CE39}
\DeclareRobustCommand{\orcidicon}{%
        \begin{tikzpicture}
        \draw[lime, fill=lime] (0,0) 
        circle [radius=0.16] 
        node[white] {{\fontfamily{qag}\selectfont \tiny ID}};
        \draw[white, fill=white] (-0.0625,0.095) 
        circle [radius=0.007];
        \end{tikzpicture}
        \hspace{-2mm}
}

\newcommand{\orcidPS}{\href{https://orcid.org/0000-0001-7044-3809}{\orcidicon}}
\newcommand{\orcidAZ}{\href{https://orcid.org/0000-0001-9509-7316}{\orcidicon}}

\newcommand{\orcidDA}{\href{https://orcid.org/0000-0002-1959-7201}{\orcidicon}}

\begin{document}
   \title{The role of magnetic field and stellar feedback in the evolution of filamentary structures in collapsing star-forming clouds}

   \author{P.~Suin
          \inst{1}\thanks{\email{paolo.suin@lam.fr}}\orcidPS
          \and D.~Arzoumanian\inst{2,3,4}\orcidDA
          \and
          A.~Zavagno \inst{1,5}\orcidAZ
        \and P.~Hennebelle\inst{6}
        }

   \institute{Aix Marseille Univ, CNRS, CNES, LAM Marseille, France 
      \and 
    Institute for Advanced Study, Kyushu University, Japan
   \and
   Department of Earth and Planetary Sciences, Faculty of Science, Kyushu University, Nishi-ku, Fukuoka 819-0395, Japan
   \and Division of Science, National Astronomical Observatory of Japan, 2-21-1 Osawa, Mitaka, Tokyo 181-8588, Japan
    \and 
   Institut Universitaire de France, 1 rue Descartes, 75005 Paris, France 
   \and
   AIM, CEA, CNRS, Université Paris-Saclay, Université Paris Diderot, Sorbonne Paris Cité, F-91191 Gif-sur-Yvette, France           
             }

   \date{Received 17/01/2025; accepted 24/04/2025}

\authorrunning{P. Suin et al.}
\titlerunning{Role of magnetic fields and stellar feedback in filaments' evolution}

  \abstract
   {Filaments are common features in molecular clouds and they play a key role in star formation. Studying their life cycle is essential to fully understand the star formation process.
   }
   {Using high-resolution magnetohydrodynamical simulations including early stellar feedback (jets and \hii regions), we aim to characterise the impact of magnetic field ($B$) and stellar feedback on the evolution of filamentary structures in star-forming clouds.}
   {We performed two numerical simulations of a collapsing $10^4\,\MSun$ cloud with different mass-to-flux ratios ($\mu=2$ and $\mu=8$). Using \disperse, we extracted the three-dimensional filamentary network and analysed its morphological and physical properties as it evolves throughout the star formation event. 
   }
   {We find that the $B$ plays a central role in shaping the density structures as the clouds evolve. With a weak field, the cloud develops a single central hub, while in strong $B$ fields the cloud maintains a sparser filamentary network. A stronger magnetisation also delays star formation, although the two runs ultimately achieved similar star formation efficiencies.
   We observed that the filaments in the simulations  follow two distinct evolutionary pathways. In the strongly magnetised case, filaments are predominantly perpendicular to $B$ lines, favouring a parallel alignment in the weak field cloud. Furthermore, while always accreting, filaments exhibit faster flows towards the hub relative to the surrounding gas. In the weakly magnetised run, the central hub dominates the dynamics, and filaments exhibit faster flows as they approach the central hub. Finally, once the expanding \hii region impacts the filaments, they align to $B$ independently of the initial configuration.
   }
   {
   Magnetic fields play a critical role in shaping the structure and dynamics of molecular clouds. Stronger magnetic fields slow the cloud's evolution and inhibit the formation of central hubs, promoting a broader filamentary network instead. 
   However, ionising feedback dominates the late-stage evolution, overriding the initial differences and dictating the final filament configuration.
   }
   

   \maketitle

\section{Introduction}\label{intro}
Filamentary structures are ubiquitous features of molecular clouds and they play a fundamental role in the star formation (SF) process. Observations from the Herschel Space Observatory have revealed that molecular clouds are highly filamentary on all scales \citep{Molinari2010, Schisano2014, Andre2014}, with filaments hosting several sites of SF \citep{Andre2010, Konyves2015, Pineda2023}. The formation and evolution of such structures are characterised by a complex interplay of physical processes, including turbulence, gravity, magnetic fields, and feedback from massive stars. 

While molecular clouds exhibit a variety of filamentary configurations, we can broadly categorise them into two types. The first type of cloud hosts many filaments arranged in an intricate web with several hubs at the filaments intersections but without a clear central convergence point. These `filamentary clouds' exhibit relatively low levels of SF activity. This occurs primarily in the hubs, which are sparsely located along the filaments and at their junctions \citep{Goldsmith2008, Lada2008}. The observed turbulence levels within these filaments are generally low to moderate, on the order of $0.2$–$1.5\,\kms$ \citep{Arzoumanian2013, Andre2014, Hacar2017, Hacar2023}. Examples of this category include the Perseus and Taurus molecular clouds and the Pipe nebula \citep{Lada2008, Panopoulou2014, Zhang2022}.

Conversely, the second type displays a single central clump where all the filaments converge in a radial pattern. These clouds are often referred to as `hub-filament systems' \citep[HFSs;][]{Myers2009}. Different from the former category, these clouds display higher column densities compared to filamentary clouds \citep{Myers2009, Hacar2018} and intense SF activity concentrated in central hubs \citep{Schneider2012, Trevino2019, Seshadri2024}. These regions often display high-velocity dispersions (a few $\kms$) and strong velocity gradients \citep{Peretto2013, Peretto2014, Hacar2018, Hwang2022}, suggesting strong converging flows powered by the central potential well \citep{Friesen2013, Trevino2019, Ma2023}. Monoceros R2 and Serpens Core are prominent examples of such configurations \citep{Roccatagliata2015, Trevino2019}. 

Some studies have proposed that these two cloud categories may represent stages in an evolutionary sequence \citep{Schneider2012, Li2014, Wang2020}. In this scenario, molecular clouds start as complex filamentary networks arising from turbulent motions and random shocks in the interstellar medium. As gravity overcomes other forces, the filamentary clouds evolve into HFSs by merging structures and through global collapse \citep{Vazquez-Semadeni2009, Vazquez-Semadeni2019}. This hierarchical collapse leads to mass accumulation in a central hub, promoting the formation of a massive cluster.
However, the situation might not be so straightforward, as magnetic field strength and orientation, turbulence levels, and external environmental effects may play crucial roles in determining the observed morphologies of molecular clouds \citep{Hennebelle2013, Inoue2016}.

The magnetic field morphology also varies between these two types of clouds. Observations have revealed that filaments generally show a bimodal alignment with the underlying magnetic field, with relative orientations that are either parallel or perpendicular to the local magnetic field direction \citep{Li2013, Palmeirim2013, plankcoll2016}. In filamentary clouds a global trend towards perpendicular alignment has been observed between the dense filaments and the field lines \citep[for example in Taurus;][]{Chapman2011, Palmeirim2013}, while in HFSs, filaments appear to align more parallel to the magnetic field lines (such as Monoceros R2 and Serpens Core; \citealt{Sugitani2010, Hwang2022}; see also \citealt{Wang2020, Wang2022, Khan2024} for other examples).

Finally, towards the end of the SF event, ionising feedback from massive stars can also play a role by heavily rearranging the parent filamentary structures. The intense flux of high-energy photons ionises the gas and disperses the surrounding cloud, reshaping the impacted structure and deeply affecting dynamics, as suggested by both observations \citep{Hester1996, Minier2013, Kruijssen2019, Zavagno2020} and simulations \citep{Walch2012, Kim2018, Zamora-Aviles2019, Suin2024}. 

Understanding the respective influence of all the actors at play is challenging in observations due to the complexity of disentangling the different mechanisms without accessing the temporal evolution of the system. In this sense, numerical simulations have become vital tools in studying the filaments' life cycle, as they allow one to follow the temporal evolution of these structures through time and  disentangling the effect of different physical processes. In fact, studies focusing on the role of the magnetic field have shown that its strength dramatically affects the properties of the arising filamentary structure \citep{Chen2016,  Zamora-Aviles2018}, while works centred on the impacting ionising radiation show how this restructures the morphology of existing filaments and leads to the formation of pillars and globules \citep{Gritschneder2010, Dale2012, Klassen2017}.

However, few studies have attempted to analyse both the growth and evolution of filaments in a realistic scenario where all the mechanisms act simultaneously. In this work, we present the evolution of a $10^4\,\MSun$ molecular cloud exposed to stellar feedback and different intensities of the ambient magnetic field. We analyse the cloud structure's evolution and the filaments' dynamics by studying the evolution of their orientation relative to the local magnetic field and the associated local velocity field.

The paper is organised as follows: Sect.~\ref{simu} presents the numerical simulations and the initial parameters used for this work. The methods are explained in Sect.~\ref{methods}, where we describe how we extracted the filaments and their properties from the simulations' datacubes. Results are presented in Sect.~\ref{res} and discussed in Sect.~\ref{discussion}. A summary and conclusions are given in Sect.~\ref{conc}.

\section{Numerical simulations} \label{simu} 
\subsection{Code and numerical parameters}
The two ideal magneto-hydrodynamic (MHD) simulations used in this study were performed with the adaptive mesh refinement (AMR) code \texttt{RAMSES} \citep{Teyssier2002}. The first of the two simulations has been presented in \citet{Verliat2022} and \citet{Suin2024} -- labelled in the two works as \texttt{HIIR+PSJ}. To this realisation, we added a new twin simulation characterised by a stronger initial magnetic field. We set the domain as a cubic box of $L=30.4\,\pc$ with periodic boundaries and an initial resolution of $128^3$ cells. When the local Jeans length becomes smaller than 40 times the cell size, the cells are split into a new level of refinement up to five times, achieving the maximum resolution of $7.4\times10^{-3}\,\pc$ $(1.5\times10^3\,\mathrm{AU})$. 
The code follows the SF by creating sink particles \citep{Bleuler2014} at a density threshold of $10^7\,\ndens$, when the Jeans length matches the smallest cell size achievable. 
Once a sink forms, it interacts with the ambient medium through the accretion from the surroundings, the ejection of protostellar jets, and emission of ionising radiation (see Sect.\,\ref{sec: stellar feedback}). 

The gas thermal behaviour was treated with a parametrised cooling function \citep{Audit2005}, which is built on the works of \citet{Wolfire1995, Wolfire2003}. It includes the principal cooling mechanisms such as CII and OI fine structure lines, electron recombination onto charged grains, and photo-electric heating polycyclic aromatic hydrocarbons (PaH) due to the external galactic UV background. Additionally, the \texttt{RAMSES-RT} \citep{Rosdahl2013} module handles the radiative cooling and heating for the ionised gas.

\subsection{Initial conditions}
The $10^4\,\MSun$ clouds were initialised as Bonnor--Ebert spheres of $15.2\,\pc$ diameter so that their number density $n$ decreases as a function of the radius as
\begin{equation}
    n(r) = \frac{n_0}{1+\left(\frac{r}{R_0}\right)^2},
    \label{eq: cloud density distribution}
\end{equation}
where $n_0=800\,\ndens$ and $R_0=2.5\,\pc$ define the central region's central density and scale. Outside the cloud, the domain is filled with a uniform density of $8\,\ndens$ up to $r=L/2$, and $1\,\ndens$ up to the box's corners. 
The initial temperature of the gas was set at the equilibrium temperature given by the parametric cooling function. This led to a temperature of $\approx 10\,$K in the denser inner region.

We initialised the turbulence with a Kolmogorov power spectrum, normalising its intensity such that the turbulent crossing time in the central region is roughly equal to its free-fall time. This set the Mach number to roughly $6.7$, corresponding to a velocity dispersion of $\approx 1.6\,\kms$ and a virial parameter $\alpha_\mathrm{vir}=2E_\mathrm{kin}/E_\mathrm{grav}\approx0.75$. These values fall within the observed range of velocity dispersions and virial parameters observed in star-forming clouds \citep{Kauffmann2013}.

Finally, the magnetic field was initialised with a uniform mass-to-flux ratio ($\mu$) across the domain oriented along the $x$ direction. In the low-magnetisation case, we set this ratio to $\mu=8$, while in the high-magnetisation run, we set it to $\mu=2$. These values result in central magnetic field strengths of $5\,\mu$G for the low-magnetisation case and $20\,\mu$G for the high-magnetisation case, which is consistent with observed magnetic field intensities in regions with a density of $800\,\ndens$ \citep[cf. with: ][]{Crutcher2010, Pattle2023}. In the central portion of the cloud, this corresponds to a plasma beta ($\beta$), the ratio between the thermal pressure and magnetic pressure, of $\approx1.24$ in the first case and $0.08$ in the second. Hereafter, we identify the two simulations as \texttt{Low\_B} and \texttt{High\_B}, respectively.

\subsection{Stellar feedback}
\label{sec: stellar feedback}
Protostellar jets and ionising radiation from massive stars are implemented in the two simulations. Once a sink exceeds $0.15\,\MSun$, at each timestep it ejects 1/3 of the accreted mass into bipolar outflows. We choose the opening angle of these ejecta to be $20^\circ$, while their speed matches 24\% of the escape velocity\footnote{We refer to \citet{Verliat2022} for a detailed description of how these jets are implemented and the choice of the jets' parameters.}. 

As for the ionising stars, a prescription had to be adopted to include them in the simulation. 
Indeed, the initial mass function arising from the simulation hardly represents the one outlined from observations \citep{Salpeter1955, Chabrier2003}. Observationally, we would expect at least one ionising star (with a mass greater than $8\,\MSun$) in a stellar population of approximately $120\,\MSun$. By the end of the simulations, the most massive star barely reaches $5\,\MSun$ despite the clouds producing more than a thousand solar masses in stars. 

This partly comes from the lack of infrared radiation from the newborn sinks, which heats the surrounding dense gas and affects the fragmentation process \citep{Krumholz2017, Hennebelle2020}. In any case, present simulation studies still struggle to derive the correct stellar mass function. \cite{Lee2018} found that much higher resolutions are generally required to reach convergence in the resulting stellar mass distribution -- of the order of tens of AU. 

Therefore, in our simulations we introduce an ionising star every time the system has converted $120\,\MSun$ of gas into sinks. Whenever the threshold is reached, the algorithm randomly draws a mass for the new star from a \citet{Salpeter1955} distribution between 8 and $120\,\MSun$. Then, the algorithm selects the sink presenting the highest accretion rate, which begins to radiate according to the parametric mass-UV luminosity relation given in \citet{Colling2018}.
To directly compare the impact of \hii regions in the two magnetisation cases, we use the same sequence of extracted stellar masses for the two simulations.

\section{Methods} \label{methods}
\subsection{Filament extraction}
\label{sec: methods_fil_extraction}

\begin{figure*}
\centering
\begin{tikzpicture}[
  mybox/.style={
    rectangle,
    draw=black,
    rounded corners,
    align=center,
    inner sep=5pt,
    minimum height=6cm
  },
  myarrow/.style={
    -latex,
    line width=1.2pt,
    shorten >=5pt,
    shorten <=5pt
  }
]

\node[mybox] (box1) {
  \includegraphics[width=0.275\linewidth]{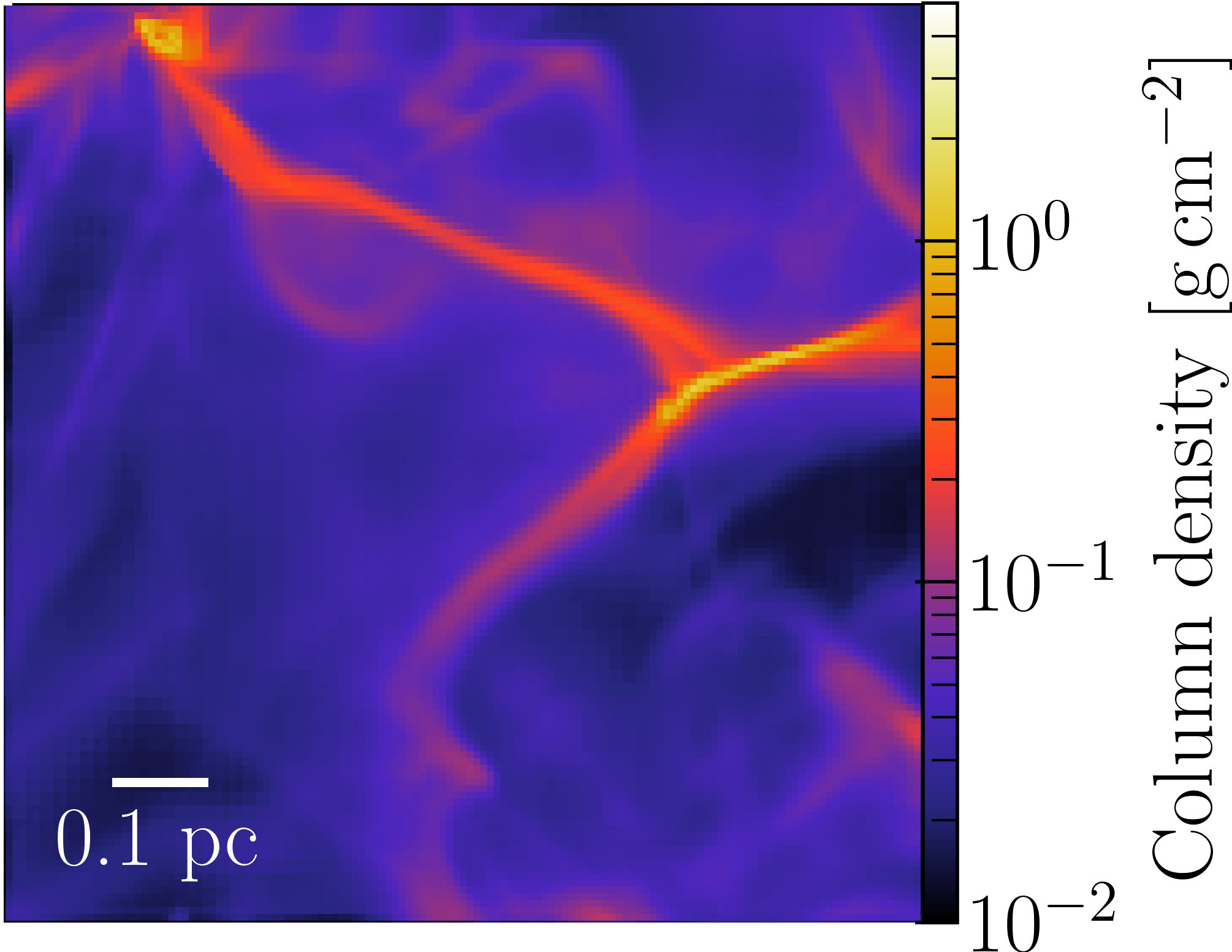} \\[0.5em]
  \textbf{Native AMR grid}\\
  $\Delta x \approx 0.007\,\mathrm{pc}$%
};

\node[mybox, right=0.05\linewidth of box1] (box2) {
      
  \includegraphics[width=0.275\linewidth]{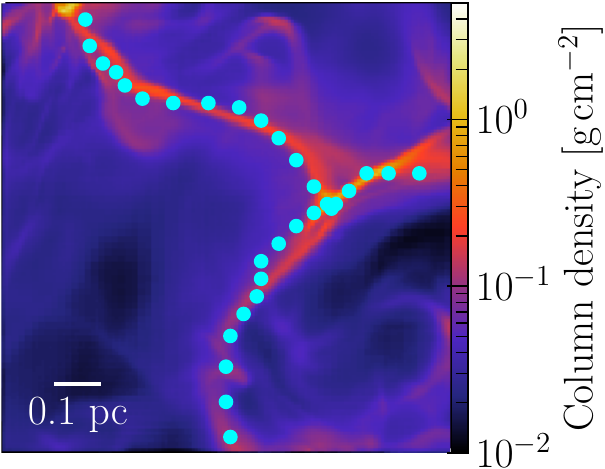} \\[0.5em]
  \textbf{Uniform grid ($128^3$)}\\
  \disperse filaments\\
  $\Delta x \approx 0.08\,\mathrm{pc}$
  
};

\node[mybox, right=0.05\linewidth of box2] (box3) {
  \includegraphics[width=0.275\linewidth]{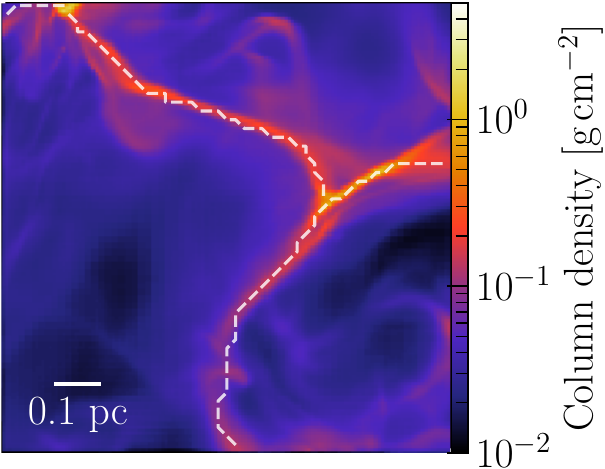} \\[0.5em]
  \textbf{Uniform grid ($512^3$)}\\
  $\Delta x \approx 0.02\,\mathrm{pc}$%
};

\draw[myarrow] (box1.east) -- (box2.west);
\draw[myarrow] (box2.east) -- (box3.west);
\end{tikzpicture}

\caption{Workflow to extract the filaments from the original data. The AMR grid (left box) is converted to a uniform grid readable by \disperse. Then, the algorithm detects the approximate locations of filament crests (middle box). Lastly, the filament extraction is refined finding the points of maximum density in the vicinity of \disperse points and connecting them (right box). \label{fig: disperse_workflow}} 

\end{figure*}

In this Section, we summarise the primary methods for extracting filaments from observational and simulation data, focusing on their application in 2D and 3D contexts. We then detail the approach used in our 3D MHD simulation datacubes employing the \disperse algorithm.

Many methods exist to extract filaments from data (see \cite{Hacar2023} and \citet{Berthelot2024} and references therein), but can be comprehensively divided into four main categories. 

1) Multi-scale decomposition methods. This class of algorithms performs a simultaneous analysis across a wide range of spatial scales using filtering or transforms which depend on a characteristic length. An example of such codes is \texttt{getsf} \citep{Men'shchikov2013, Men'shchikov2021}. Although effective in capturing structures across various scales, these codes can be computationally intensive, and require fine-tuning of parameters to distinguish between true filamentary structures and noise.

2) Non-local methods. These codes focus the analysis on more confined scales and are mainly based on template-matching methods \citep{Juvela2016}. These algorithms, such as \texttt{filfinder} \citep{Koch2015}, benefit from the application of the Rolling Hough Transform \citep{Clark2014}, which returns a measure of the linearity within a given region. They then evaluate how well the data matches specific patterns at a given scale, which is simply a line with a given width and orientation when looking for filaments. Although effective in detecting filaments with specific, well-defined shapes, and have been widely applied on Herschel column density maps \citep{Koch2015, Liu2018, Gu2024, Zhang2024}, these codes may struggle with more complex and irregular filamentary structures, exhibiting substantial variations in density and contrast. Moreover, they are hardly extended to 3D datacubes -- as is the case in our simulations -- due to the increased complexity of applying template-matching methods to volumetric data, where the additional degree of freedom makes the identification process more intricate.

3) Statistical and machine learning-based methods. Machine learning techniques have recently been explored for filament extraction \citep{Alina2022, Zavagno2023, Berthelot2024}. These methods rely on training datasets to guide the detection of filaments. The advantage of these methods lies in their flexibility and generalisation capabilities. However, they rely heavily on the availability of high-quality training data, and the results can sometimes lack the physical interpretability of more classical methods. While machine learning approaches are promising in terms of computational efficiency, they may not yet be mature enough to replace more established methods, especially when training data are scarce or highly variable.

4) Local derivative-based techniques. These methods focus on computing spatial derivatives at the pixel level, identifying local variations in intensity or column density. They can be either first-order, analysing the density or intensity gradients to detect the orientation of elongated structures \citep{Soler2013, plankcoll2016}, or second-order, using the Hessian matrix to derive the local morphology \citet{Schisano2014}. 
\disperse \citep{Sousbie2011} belongs to this category. Once the density field's critical points (minima, maxima and saddle points) are identified, the code constructs the Morse--Scale complex of the domain to extract its topological features. 
Originally developed for analysing the cosmic web \citep{Sousbie2013}, this code has been adopted in astrophysical research too, both observational studies \citep{Arzoumanian2011, Panopoulou2014} and numerical simulations \citep{Smith2014, Klassen2017}. 

While local methods can be sensitive to random fluctuations (structures created by noise fluctuations), they are easily applicable to three-dimensional datasets and ensure that the extracted structures (filaments in our case) are real topological features. Thus, although this approach may lead to a higher number of false positives, it guarantees a very low rate of false negatives, making these methods particularly suitable for the highly variable conditions of our 3D MHD simulations.

Moreover, \disperse mitigates the sensitivity to noisy structures by computing the `persistence' of the extracted features. This measure is defined as the density difference between the two extrema of the structures (the critical points characterising the filaments). By definition, noisy features tend to have low-persistence values. Therefore, upon applying a threshold, the algorithm cancels the structures having lower persistence and reconnects the correspondent Morse--Scale complex, preserving the underlying topology.

This threshold is indeed the only user-defined parameter adopted by the code. Since we are interested in following the properties of star-forming filaments, we adopted a relatively high threshold of $10^{-19}\,\dens$ ($4.2\,\times 10^4\,\ndens$), comparable to the critical density for isothermal self-gravitating filaments \citep[$\approx 1.5\times 10^{-19}\,\dens$ assuming a filament width of $0.1\,\pc$;][]{Andre2014}.
The filaments retrieved with this threshold are likely already well-developed. Lowering it would increase the number of detected filaments and potentially capture younger structures. However, it would also introduce more transient and noise-driven features, especially at later stages of the SF event (see also Appendix A in \citealt{Arzoumanian2019} and \citealt{Dhandha2024}). In Appendix~\ref{sec: appendix thr}, we examine the impact of a lower persistence threshold, confirming that the overall properties and trends remain consistent. 

Since the code cannot directly read the AMR grid of \texttt{RAMSES} outputs, we performed some preliminary steps to set up the analysis, as schematically shown in Fig.~\ref{fig: disperse_workflow}. Due to the high memory usage of \disperse, we focused on the central $10\,\pc$ of the simulation domain -- where the SF is occurring -- and mapped each output at various simulation epochs onto two uniform grids of $128^3$ and $512^3$ cells.
Then, we applied \disperse on the first coarse datacube to extract the filaments skeleton (middle box). Finally, we transposed this skeleton onto the finer grid. For each filament point, we find the maximum-density location within a cube of $4^3$ refined cells centred in its original position. Connecting these points reconstructs the final filament skeleton, as shown in the right box. 

To reduce the impact of noisy structures on the analysis, such as transient over-densities or compression produced by stellar ejecta, we excluded filaments shorter than $0.2\,\pc$. In \simA, where the evolution leads to the formation of a compact central hub (see Sect.~\ref{sec: cloud evolution}), we also masked filament points closer than $0.2\,\pc$ from the minimum of the gravitational potential well. In this region, the high-velocity gradients and the intense SF, make it challenging for \disperse to identify and trace true structures accurately.

\subsection{Extraction of filament properties}
Understanding the local physical properties of filaments provides crucial insight into their dynamics and evolution. Key characteristics such as magnetic field, velocity field, and density are critical for determining how filaments interact with their surrounding environment. We focus on the regions surrounding the filaments' crests to extract these properties. While global properties such as filament length and total mass can vary significantly due to large-scale morphological changes and depend on the code setup\footnote{The code contains additional parameters to handle the division of the extracted skeleton into single filaments. These include for example the maximum bending angle at which a filament is split into two, and the minimum persistence difference to merge two adjacent filaments. Since our analysis focuses on the statistical properties of the entire filamentary network rather than on individual filament properties, we configure these parameters to maximise merging (a large angle of 90$^\circ$ and a null persistence threshold for merging). This minimises the impact of the segmentation choices on the study. In general, the removal of filaments shorter than $0.2\,\pc$ filaments impacted the total length by less than $1\%$.}, the overall skeleton structure remains consistent. This ensures that, despite potential variations in the way the algorithm separates filaments, the underlying structure of the filamentary network is preserved. Therefore, by analysing the local properties at the pixel level with a statistical approach, we can achieve reliable measures of the influence of various physical processes on filament structure and behaviour.

First, we calculate the tangent direction at each point along the filament, performing a linear regression on a $0.15\,\pc$ segment centred on the reference cell. Then, we estimate the filament properties averaging over all the cells contained in a cylinder of $0.1\,\pc$ diameter and $0.1\,\pc$ height, aligned with the recovered filament’s axis and centred in the crest cell. We stress that the choice of the radius is not based on the conventional filament width \citep{Arzoumanian2011, Andre2014}, but rather on the physical resolution achieved by our simulations, generally assumed of the order of 10 resolution elements ($\approx 0.07\,\pc$ in our simulations). By averaging the properties over a larger volume, we ensure accurate measurements of the local medium conditions in the vicinity of the filaments crest. 
To reduce the impact of the discrete nature of the grid on our analysis, we subdivided each cell into $4^3$ smaller subcells, each inheriting the same physical properties as the original cell. This approach refines the averaging of quantities within the cylindrical region, allowing even partially included cells to contribute to the overall result.

To understand how the structure influences the behaviour of the surrounding gas, we also computed filament properties at varying distances from the spine. We averaged over cylindrical shells of $\approx0.04\,\pc$ thickness around the crest, to analyse the radial dependence of the various quantities up to radii of $0.7\,\pc$ from the crest. Throughout the paper, we consider shells with radii greater than $0.2\,\pc$ as representative of the gas properties outside the filament, while shells with radii smaller than $0.05,\pc$ are used to characterise the gas properties within the filament.

To fully capture the dynamics and evolution of the structures throughout the SF event, we extracted in both simulations the filamentary network at various stages of the evolution of the cloud, from the moment the first stars were born up to the final dispersion of the gas by the expanding \hii region.

In Appendix~\ref{sec: appendix before SF}, instead, we briefly study the properties of the skeleton prior to the onset of SF. We find a certain resemblance between the properties extracted at earlier phases and the ones after the formation of the first star. However, our primary focus is on the dynamical evolution and growth of filaments during the star-forming and feedback-dominated phases rather than their early formation. While some properties of early-stage filaments can be linked to their formation mechanisms through comparisons with analytical models and observations, and is in part justified with Appendix~\ref{sec: appendix before SF}, a detailed study of their formation phase is beyond the scope of this paper. We refer to works such as \citet{Arzoumanian2018} and \citet{Bonne2020} discussing the filament formation process from observations and \citet{Gomez2014}, \citet{Soler2017}, or \citet{Abe2024} for analytical and numerical explorations of filament formation models, and the recent review from \citet{Pineda2023}.

\section{Results} \label{res}
\begin{figure*}[h!]
    \centering
    \includegraphics[width=0.227\linewidth]{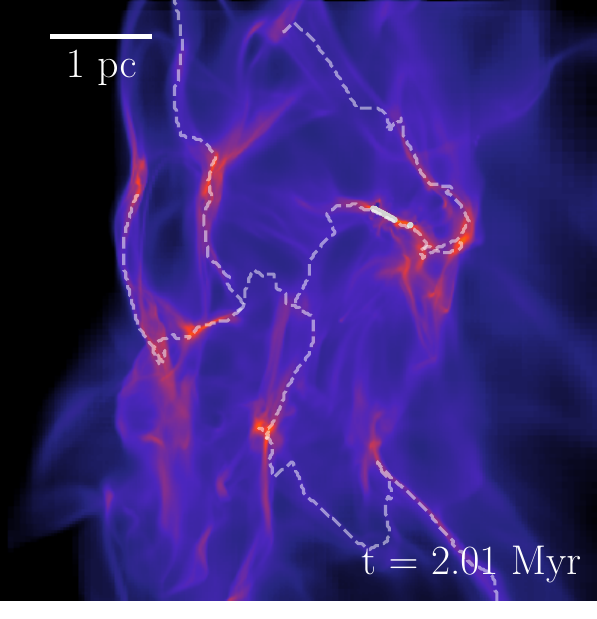}
    \hfill
    \includegraphics[width=0.227\linewidth]{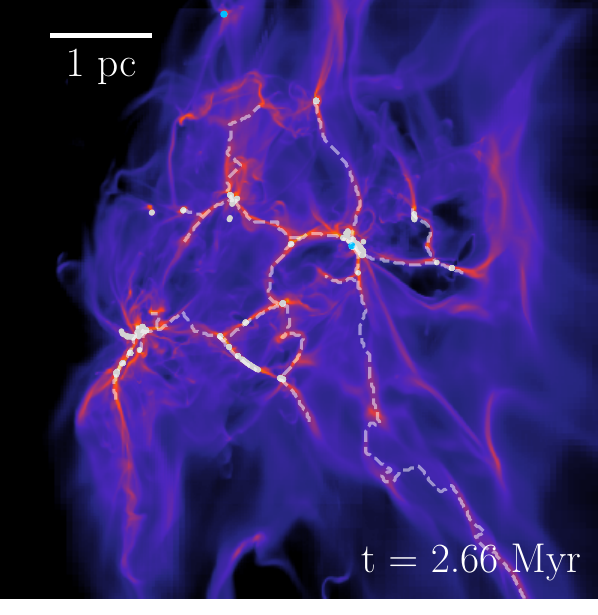}\hfill
   \includegraphics[width=0.227\linewidth]{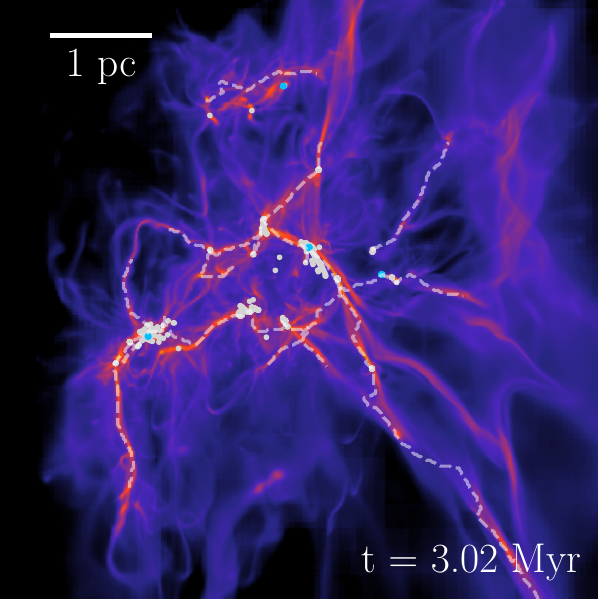}
    \raisebox{-0.02cm}{
    \includegraphics[width=0.29\linewidth]{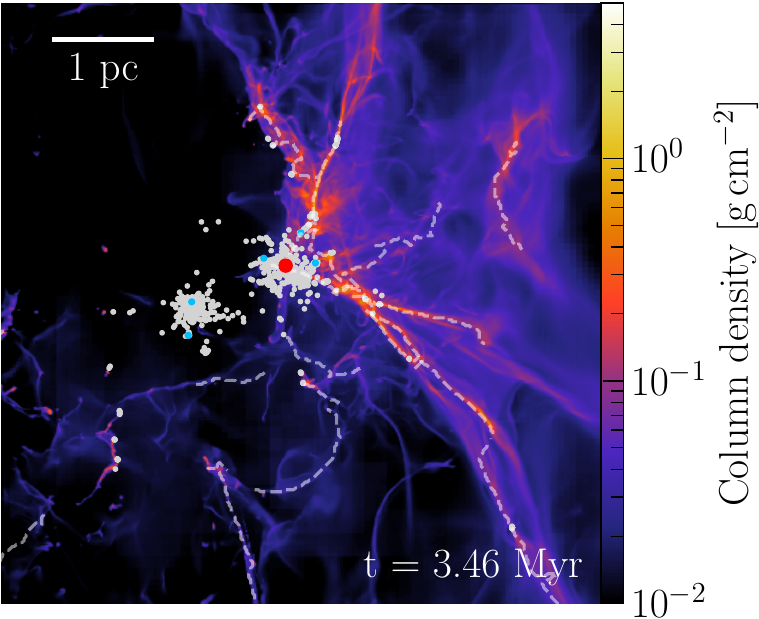}}
    \\
    \includegraphics[width=0.227\linewidth]{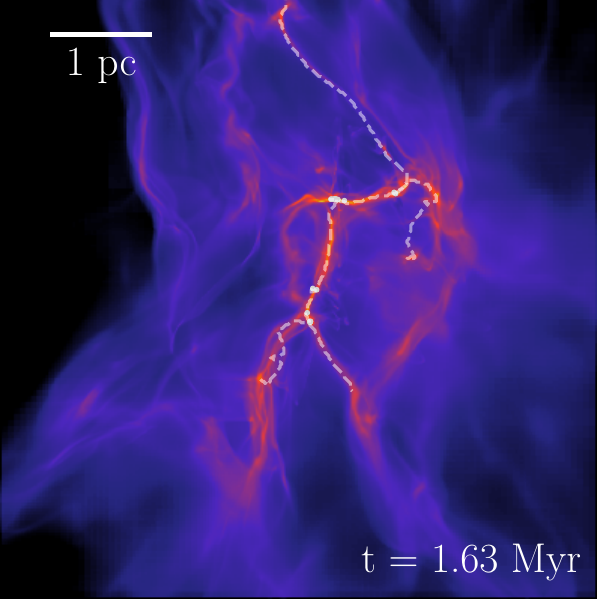}
     \hfill
    \includegraphics[width=0.227\linewidth]{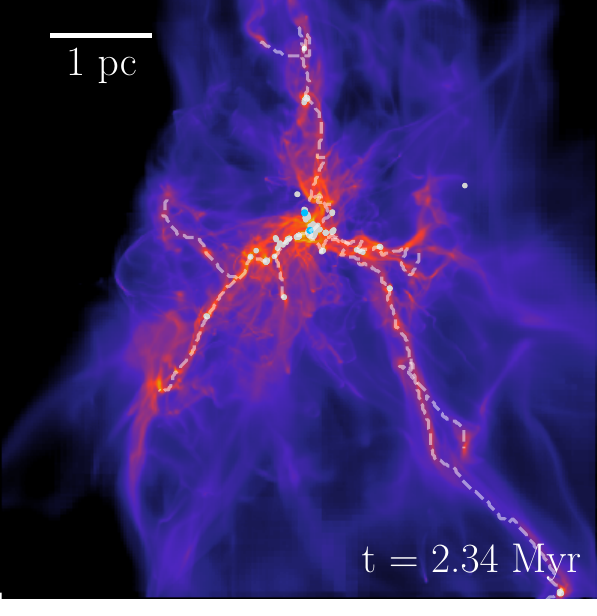}\hfill
    \includegraphics[width=0.227\linewidth]{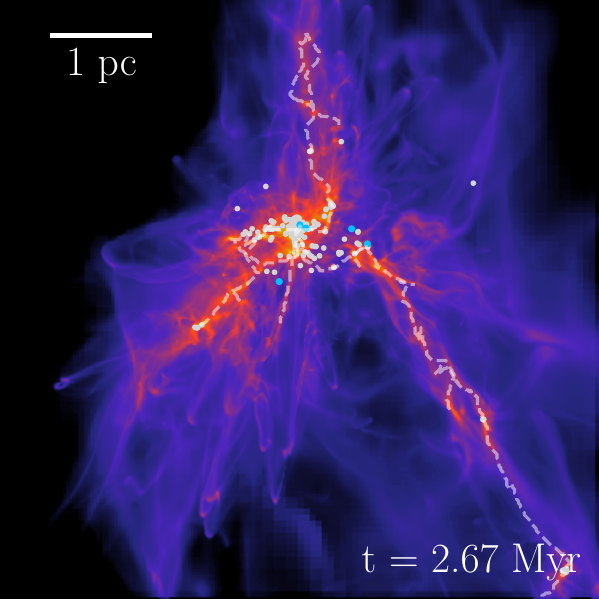}\hfill
    \raisebox{-0.02cm}{\includegraphics[width=0.29\linewidth]{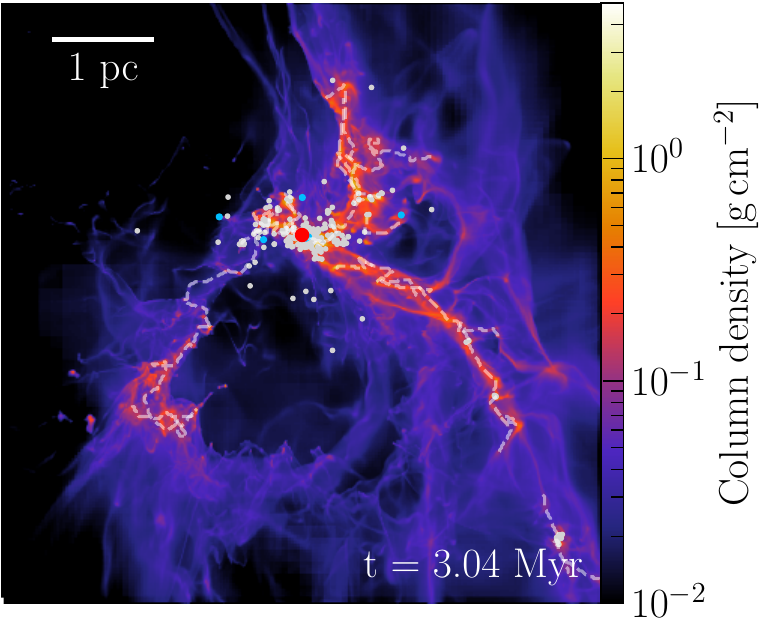}}

    \caption{
    Evolution of the same cloud in the high- (\simBB, top) and low-magnetisation case (\simA, bottom). The first three columns display the column density projections of the cloud at equal $\SFE$ (from left to right: 0.1\%, 3\% and 6\%). The last column shows how the giant \hii region reshapes the environment, showing the cloud after $0.3\,\Myr$ from the birth of the $100\,\MSun$ star. White dots mark the position of the sink particles, while blue dots highlight the ionising stars within the domain. The red dot represents the 100$\,\MSun$ star from which the main \hii originates. The grey dashed lines mark the filamentary structures retrieved by \disperse. These snapshots display the role of the $B$ field in shaping the cloud structure (single hub vs sparse filamentary network) and influencing the timescale of the SF event.
    \label{fig: cloud_simulation} 
    }
\end{figure*}

In this Section, we first describe the two clouds' global evolution and analyse the morphology of their filamentary structures. Then, we describe the dynamical evolution of the filaments' properties. We first focus on the filaments' three-dimensional orientation relative to the magnetic and the velocity field. Finally, we analyse the gas dynamics around the filaments. 

\subsection{Cloud evolution under two different magnetisations}
\label{sec: cloud evolution}
Before discussing the evolution of the filaments' dynamical properties throughout the SF process, we consider the overall evolution of the simulated cloud's structure and morphology. Due to the two different initial magnetisation levels, the cloud's behaviour differs significantly, as do the resulting properties of the filaments it hosts. Taking a broad view of the large-scale cloud evolution helps one better contextualise the changes of filament properties, which we explore in the following sections.

The importance of the magnetic field is visible with a quick look at Fig.~\ref{fig: cloud_simulation}. This shows panels to compare the structure of the two simulations (\simBB on the top row, \simA on the bottom), along with the recovered filaments by \disperse (dashed white lines), at different SF efficiencies (SFEs), defined as:
\begin{equation}
    \mathrm{SFE} = \frac{M_\mathrm{sink}}{M_\mathrm{gas} + M_\mathrm{sink}},
\end{equation}
where $M_\mathrm{sink}$ and $M_\mathrm{gas}$ are the total mass of sinks and gas in the simulation domain, respectively. The snapshots are selected to represent key moments of the cloud evolution: the beginning of the SF (first column, at a SFE of 0.1\%), the formation of the main hub in \simA (second column, $\SFE=3\%$), the moment before the onset of the \hii region that eventually disperses the cloud (third column, $\SFE=6\%$) and finally at 0.3\,Myr after the birth of the $100\,\MSun$ star (fourth column), when the ionised bubble is well developed and has restructured the surrounding environment with its expansion. By this time, the clouds in \simBB and \simA reached a SFE of $9.6\%$ and $9.4\%$, respectively. Eventually, the SF event shuts down almost entirely in both runs. The final time considered in the two simulations is $4.16\,\Myr$ in the high-magnetisation case and $3.98\,\Myr$ for the low-magnetisation one, as the SF rate drops below $100\,\MSun\,\Myr^{-1}$.

Initially, the injected turbulence promotes the formation of dense regions via shock compression. These act as seeds for the onset of SF in the cloud \citep{Klessen2000, MacLow2004, Hennebelle2013}. However, in \simBB, the stronger magnetic field provides additional support against gravity, delaying the onset of SF. Indeed, looking at the snapshots' times displayed on the plots in Fig.~\ref{fig: cloud_simulation}, one can see how the SF proceeds more slowly, with the SF event beginning roughly $0.4\,\Myr$ later than in \simA run. The magnetic field also contributes to forming a broader, more diffuse filament network in \simBB compared to \simA. In fact, the SF occurs simultaneously at multiple locations, forming small star clusters which then merge hierarchically as turbulence dissipates and gravity overcomes the other forces (first and second columns). By $3.02\,\Myr$ (third column), four main clusters are still distinguishable in the cloud, along with a few sub-clusters distributed in the surroundings. As the clusters and the connected filaments continue to join together, at $\approx 3.1\,\Myr$ the cloud achieves a certain resemblance with observed HFS, although preserving two separated main clusters still visible at $3.46\,\Myr$ (last panel). We note that such configuration of HFSs with multiple clusters can also be found in real star-forming cloud \citep[e.g.][]{Doi2020, Fiorellino2021}.

Conversely, the \simA cloud, with weaker magnetic support, develops a more compact filament distribution from the early stages. By $2.1\,\Myr$, the central filaments merge to form an HFS (second column), with a main cluster dominating the SF and the density structures aligned preferentially along a radial axis. The resulting morphology visually resembles observed HFSs, such as Monoceros R2 and the Cocoon nebula \citep{Trevino2019, Wang2020}, with long filaments extending from the central clump and feeding material onto it. Aside from a few clumps along the filaments, the central cluster dominates the SF process. 

Then, as SF progresses, ionising stars form in both simulations. The morphology of the clouds leads to distinct differences in how \hii regions impact the clouds. 
In \simBB, the stronger magnetic field keeps the cloud more diffuse, allowing the ionised bubbles to expand more freely and to more easily reshape the environment, although the impact of the first ionising stars (with masses smaller than $20\,\MSun$) is still negligible compared to the expansion of the bubble generated by the $100\,\MSun$ star (in this case forming at $3.14\,\Myr$). In \simA, instead, the compact cloud structure limits the expansion of the ionised bubbles, which do not significantly affect the global morphology until the formation of the $100\,\MSun$ star at $2.73\,\Myr$ (SFE = 7.2\%). 

Ultimately, the UV radiation from the most massive star disperses both clouds, halting the SF. At the end of the simulations ($\approx4\,\Myr$ in the strongly magnetised case and $\approx3.6\,\Myr$ in weakly magnetised one), the two clouds exhibit similar SFEs, roughly of the order of 12\% in \simBB and 13\% in \simA.

\subsection{Global structure}
\label{sec: cloud structure}
\begin{figure*}[t]
    \centering
    \includegraphics[width=1\linewidth]{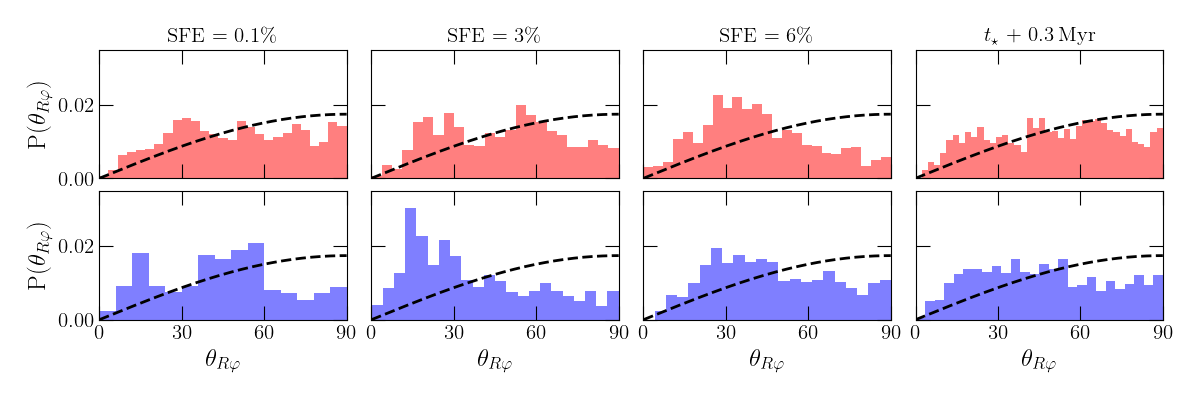}
    \caption{
    Distribution of the relative angle between the filament crests and the radial direction from the centre of the cloud in simulations \simBB (top row) and \simA (bottom row). The black dashed lines mark the random distribution for the relative orientation of two independent vectors in 3D.
    The distributions evolve similarly for both cases, but \simA filaments display a more pronounced preference towards radial alignment from early stages due to the strong influence of the central potential well. As \simBB develops a central hub (at $\SFE=6\%$), a similar radial configuration arises.
\label{fig: radial angle}}
\end{figure*}
As shown in the snapshots in Fig.~\ref{fig: cloud_simulation}, both simulations develop an overall HFS morphology (although in \simBB this process is much slower), with filaments forming and converging towards a hub at the centre of the cloud. To quantify this evolution of this overall structure and compare the differences between \simBB and \simA, we calculate the local orientation of the filaments with respect to the radial direction from the potential well of the cloud.

For each given snapshot, we first locate the position of the potential minimum in the domain, accounting also for that generated by sink particles, using a softening length of four resolution elements (default \texttt{RAMSES} value). Then, we compute the relative 3D angle between the radial direction from that minimum and the local filament direction (hereafter denoted by $\varphi$), $\theta_{R\varphi}$. We plot in Fig.~\ref{fig: radial angle} the distributions of the extracted set of angles for the same snapshots displayed in Fig.~\ref{fig: cloud_simulation} for \simBB\ (top row, red) and \simA\ (bottom row, blue), measured in each cell belonging to the identified filaments (Sect.~\ref{sec: methods_fil_extraction}). We preserve the same colour code for the simulations throughout the paper to make the graphs more immediate. 

Fig.~\ref{fig: radial angle} indicates that initially the filaments forming in the strongly magnetised cloud display a distribution closer to the expected random one\footnote{For three-dimensional vectors, this is $\propto\sin(\theta_{R\varphi})$.} (black dashed lines), although a general preference towards smaller angles can be detected. As the cloud evolves, the distribution of \simBB moves towards more parallel configurations, marking the slow overcoming of gravity on the other forces and the progressive drift towards the formation of a central hub. Filaments in \simA, instead, show a preference for alignment with the radial direction from the first panel ($\SFE=0.1\%$). This yields a peak at smaller $\theta_{R\varphi}$ values, suggesting converging flows of the filaments towards the minimum of the potential well. 
\begin{figure*}[t]
    \centering
    \includegraphics[width=1\linewidth]{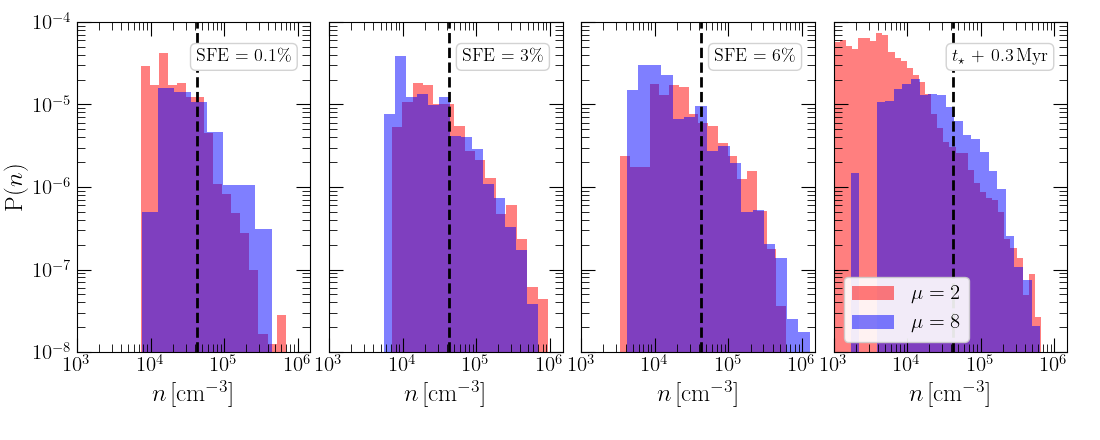}
    \caption{
    Distribution of the number density $n$ of the filament crests in \simBB (red) and \simA (blue). The black dashed line marks the persistence threshold adopted. Filaments in \simBB are initially less dense, and require more time to achieve a distribution similar to \simA. After the onset of the \hii region, the sparser morphology of \simBB cloud causes the distribution to spread towards lower densities compared to \simA.
\label{fig: densities}}
\end{figure*}

After the eruption of the \hii region, the expanding shell pushes the density structures and realigns them into more perpendicular orientations, shifting the distributions to higher $\theta_{R\varphi}$. Despite this, the distribution remains skewed towards parallel alignment compared to the random configuration.  This is partly because the $100\,\MSun$ star responsible for the \hii region's formation is not located at the exact centre of the cloud, causing the bubble to expand asymmetrically\footnote{From the panels in the last column of Fig.~\ref{fig: cloud_simulation} it may appear that the main ionising star forms inside of the cluster.  We note, however, that in both simulations the star forms in an infalling filament towards the left of the central hub (as one may evince from the bubble visible in the $\mu=2$ case, top right panel). Only after 0.2\,Myr the cluster captures the star.}. Moreover, pillar-like structures forming in the simulations extend their tail parallel to the direction of the ionising radiation, contributing to more parallel alignment. 

The varying evolution of the filament networks is also evident in their density distributions. Figure~\ref{fig: densities} shows the probability distribution function of the crests' number density, $n$, in the two simulations (\simBB in red and \simA blue), computed averaging over $0.05\,\pc$ radius cylinders (Sect.~\ref{sec: methods_fil_extraction}). The black dashed line marks the persistence threshold adopted\footnote{We note that the persistence threshold is not a precise indicator of the minimum density at which the sample becomes incomplete. Since it measures the density difference between the two critical points defining a filament, it only filters out structures whose maximum density is near or below the threshold, rather than constraining the lowest density along the filament.} ($4.2\,\times 10^4\,\ndens$, see Sect.~\ref{sec: methods_fil_extraction}). The presence of the magnetic field contrasts the action of gravity, and as a result the filaments of \simBB are less dense at the beginning of SF event (leftmost panel at $\SFE=0.1\%$). 
 
While the peak of the filament density distribution in Fig.~\ref{fig: densities} is likely influenced by the selected \disperse threshold, the difference in the distributions' high-density tail gives us hints on the evolution of the filamentary structure. At the earliest phase (first panel), we can see how the distribution arising from \simBB is shifted towards smaller densities, manifesting the effects of the stronger field on the first structures forming. As time progresses ($\SFE=3\%, 6\%$ panels) the two distributions tend to even out, as gravity eventually overcomes the magnetic support in \simBB and becomes the primary driver. We outline, however, that filaments in \simBB reach a similar density distribution to those in \simA at the same $\SFE$ of 3\%, but at a later stage in the simulation, as the corresponding snapshot is delayed by $0.3\,\Myr$ (see Fig.~\ref{fig: cloud_simulation}).

Afterwards, the onset of the \hii region affects each simulation differently. In \simBB, where the magnetic field preserves a sparser cloud structure, the filaments become extremely diffuse and exhibit a steepened slope, while in \simA, the dense structures are further compressed, shifting the peak of the distribution at higher $n$. This divergence highlights the complex interplay between magnetic fields, gravity, and ionising radiation in regulating the density and morphology of star-forming filaments.

\subsection{Filament evolution}
The evolution of the cloud in \simBB and \simA highlights the influence of the magnetic field on the filamentary structures. These differences show up not only in the overall cloud structure but also in the way filaments form, evolve, and interact with their environment. In the following subsections, we examine the evolution of the filaments' properties as they develop throughout the SF event.

Since the magnetic field ($B$) is the only variable factor between the two simulations, it is crucial to isolate its role in shaping filament dynamics. 
Thus, the first step is to understand how the magnetic field interacts with the velocity field, shaping both the gas motion and the evolution of the structures. 
Therefore, we begin by examining the relative orientation between the magnetic field and the velocity field with the filamentary structures, as this approach allows to analyse the surrounding dynamics within a well-contextualised framework.
Afterwards, we move on to study the filaments' dynamics, investigating how the magnetisation affects the surrounding gas motion and consequent accretion processes, and how these properties can be used to characterise the different configurations of the cloud (HFS or filament network) developed in the two simulations.

\begin{figure*}[t]
    \centering
    \includegraphics[width=1\linewidth]{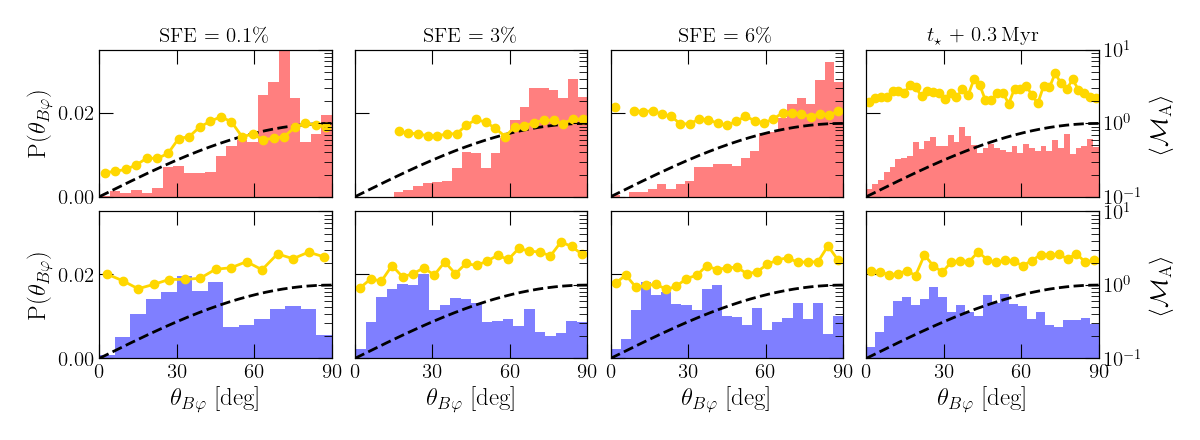}
    \caption{
    Evolution of the distribution of the relative angle between the magnetic field and the filament ($\theta_{B\varphi}$) in \simBB\ (top) and \simA\ (bottom). The black dashed lines mark the random distribution for the relative orientation of two independent vectors in 3D. The yellow dotted lines mark the average Alfvén Mach number $\mathcal{M}_\mathrm{A}$ for filament portions at a given $\theta_{B\varphi}$ (scale on the right axis). With a strong magnetic field, filaments align perpendicularly to $B$, while a weak magnetic field stimulates parallel configurations. The positive dependence of $\mathcal{M}_\mathrm{A}$ on $\theta_{B\varphi}$ and the angle distribution suggests that filament orientation relative to $B$ is governed by alfvénic modes in strong fields and by turbulent flows in weak fields.
\label{fig: angle Bphi SFEHist}}
\end{figure*}
\subsubsection{Filament orientation relative to the magnetic field}
\label{sec: orientation of filament}
The relative orientation between the magnetic field, $B$, and the star-forming filaments is an important diagnostic tool to understand the role of magnetic fields and \hii regions in the formation and evolution of filaments. Since the magnetic field introduces an anisotropy in the gas motion, studying this orientation provides valuable insights into how it influences both the filament formation and the surrounding gas dynamics. 
In the following, we examine how this orientation manifests in our simulations and how it evolves under different magnetic environments.

Figure~\ref{fig: angle Bphi SFEHist} shows the normalised distribution of the relative angle between $B$ and the filament crests ($\theta_{B\varphi}$), in \simBB\ (top row, red) and \simA (bottom row, blue). The data demonstrate that the strength of $B$ significantly influences the evolution and orientation of the filaments. In \simBB filaments appear to form mainly perpendicular to the magnetic field (leftmost panel, see also Appendix~\ref{sec: appendix before SF}). This perpendicular configuration is not only preserved during the SF event, but also enhanced, up to the onset of the main \hii region at $\SFE \approx 7\%$.  When compared to the expected random distribution for three-dimensional vectors (black dashed line), it becomes evident that underlying physical processes drive this trend. In fact, applying a Kolmogorov--Smirnov (KS) test, we confirm the distribution of $\theta_{B\varphi}$ to be different from the random one\footnote{The test has been applied to all the snapshots, in both simulations. In all cases, the $p$-value resulted in orders of magnitude smaller than the conventional threshold of 0.05. For this reason, the exact results of the other KS--test have not been reported in the text. \label{footnote: KS test}}. 
Conversely, in \simA, the distribution shows a preference for alignment between the magnetic field and the filaments since their formation, as indicated by the excess in the parallel portion of the graph. 

Such behaviours align with the analytical description proposed in \citet{Soler2017}. In environments with strong magnetic fields, as in \simBB, the field lines resist bending, guiding the gas to flow along them and leading to the formation of filaments perpendicular to the magnetic field. In weaker magnetic field environments, like in \simA, the magnetic field is unable to dominate the gas dynamics in the turbulent flow. The gravitational collapse drags the magnetic field resulting in a preferentially parallel configuration.
In addition, according to the model, both perpendicular and parallel orientations act as attractors, and the system will tend towards one or the other depending on the initial magnetic configuration. This explains the increasing perpendicularity observed in \simBB as the simulation progresses.

Finally, an interesting feature comes from the last panel ($t_\star + 0.3\,\Myr$), illustrating the situation after 0.3\,Myr from the onset of the \hii region. As visible from Figs.~\ref{fig: cloud_simulation}-\ref{fig: radial angle}, by this time the ionising flux and the expansion of the hot bubble have already restructured the cloud and dominate the system dynamics in both simulations. Here, we see a convergence between the distributions of $\theta_{B\varphi}$ extracted from the two simulations, with both leaning towards the parallel configuration. This shift is real (see footnote~\ref{footnote: KS test}) and is especially interesting in \simBB, where it reverses the initial trend towards perpendicularity.  

Multiple physical mechanisms could be responsible for this switch to parallelism. One possible interpretation is that the temperature rise caused by ionisation increases the sound speed, lowering the $\beta$ parameter and pushing the system towards the weak magnetic field regime described by \citet{Soler2017}. In this way, the systems return to a condition similar to the first phases of \simA. However, since the density structures have already formed in the environment, it is unclear whether the theory can still be applied. Another cause could be a sort of projection effect caused by the shock. Expanding, the ionised bubble forces both $B$ and the density structure onto the two-dimensional edge of its front. This decrease in dimensions might contribute to the observed shift towards more parallel configuration\footnote{Take as an example the random distribution that goes from $\propto \sin(\theta_{B\varphi})$ in the three-dimensional case to the uniform distribution in 2D, increasing the probability of observing more parallel configurations.}. This matches with observations of \hii region shells reporting a magnetic field aligned with the edges of the front \citep{Arzoumanian2021, Tahani2023, Bij2024}.

\begin{figure}[t]
    \centering
    \includegraphics[width=0.34\linewidth]{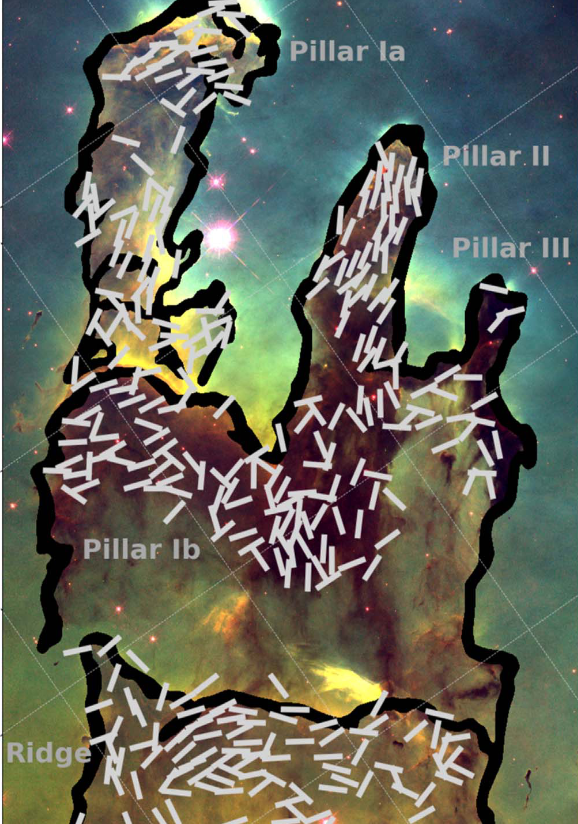}
        \hfill
    \includegraphics[width=0.63\linewidth]{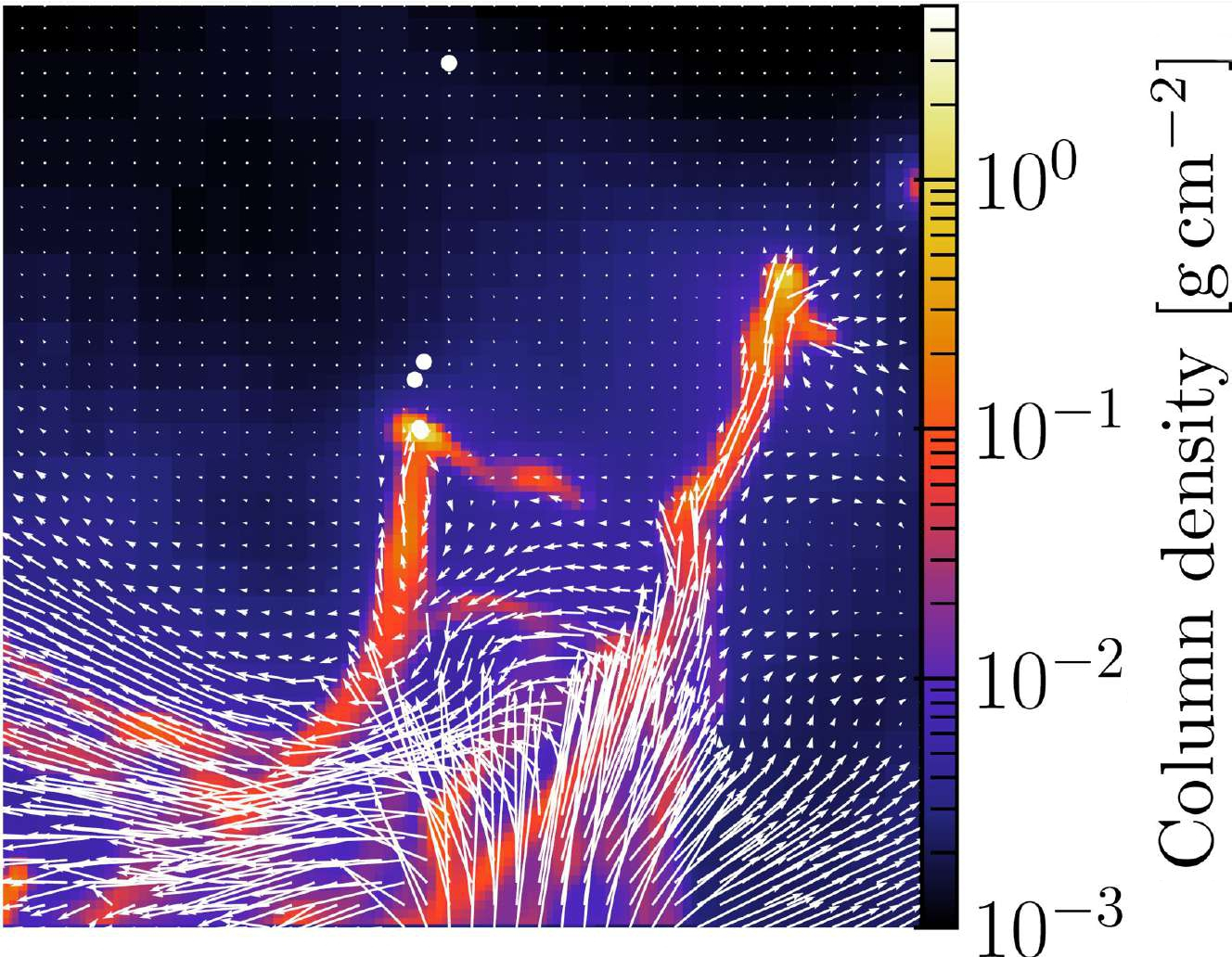}
    \caption{
    Left: Observed magnetic field in the Pillars of Creation from \citet{Pattle2018}. Right: Projection zoomed in on two pillars forming in \simBB at $3.34\,\Myr$ of $1\,\pc\times1\,\pc$ with the direction of the magnetic field overplotted. The visual comparison identifies three zones: the head, where the magnetic field remains anchored to the gas; the body, where it is dragged and aligns with the filament or pillars; and the base, where the expansion compresses $B$ onto the shell of the \hii region.
\label{fig: pillars}}
\end{figure}

Refining the picture, such a description is also compatible with recent observations of pillars protruding towards the centre of the \hii regions \citep{Hwang2023}. These show the magnetic field aligned with the density structure within the column of the pillar, and directed perpendicularly to the direction of the ionising flux within the head of the pillar. As an expanding bubble impacts a clump, it aligns the magnetic field to it. Then, the clump acts as an anchor and the magnetic field bends around it as the front passes ahead \citep{Pattle2018}. The resulting pillar will preserve the direction of the magnetic field in the head but will end up parallel to it in the newly formed tail. This scenario is supported by recent observations conducted in active star-forming regions \citep{Pattle2018, Bij2024}. In Fig.\,\ref{fig: pillars}, we propose a visual comparison between the magnetic field lines in the Pillars of Creation, as observed by \citet{Pattle2018}, and a set of pillars forming in \simBB simulation.

\subsubsection{Impact of relative orientation on physical properties}
To give more insights into the underlying physical processes, we also analysed the coupling between the filament orientation relative to $B$ and the local relative importance of the magnetic field in the local dynamics, quantified by the Alfvén Mach number. This is defined as 
\begin{equation}
    \machav = \frac{\sigma}{c_\mathrm{A}},
\end{equation}
where $c_\mathrm{A}$ is the local Alfvén speed ($c_\mathrm{A}=\sqrt{\frac{B^2}{4\,\pi\,\rho}}$, $\rho$ the local volume density) and $\sigma$ is the local velocity dispersion of the gas\footnote{The volume considered for each cell corresponds to the cylinder defined in Sect.~\ref{sec: methods_fil_extraction}, with diameter $0.1\,\pc$ and same height.}. The yellow line in Fig.~\ref{fig: angle Bphi SFEHist} shows its dependence on the relative inclination angle (yellow line with the axis on the right), obtained averaging $\machav$ of filaments' cells belonging to the same $\theta_{B\varphi}$ bin. In the earliest stages (first column), the $\machav$ is lower within filament crests aligned with the magnetic field. Indeed, in sub-alfvénic turbulence, slow modes become important in the overall flow. These are perpendicular to the magnetic field and thus cause the density structures to form aligned with it. Instead, if the turbulence is super-alfvénic, the fastest mode dominates, creating density structures perpendicular to $B$ \citep{Chen2016, Soler2017}. 

Looking at the evolution of the $\machav$--$\theta_{B\varphi}$ relation in \simBB (first row), we see that the two variables show a strong positive correlation during the formation phase. $\machav$ reaches values  as low as 0.2 at low $\theta_{B\varphi}$, and $\machav \approx 1$ as the relative angle increases. This means that filaments with parallel orientation to the magnetic field are mostly found in regions where the action of the magnetic field is stronger. However, as the cloud evolves, this dependence weakens, settling on a value of roughly 1 at every relative orientation. 
On the contrary, in \simA the relation between $\machav$ with $\theta_{B\varphi}$ does not evolve significantly in the low-magnetisation cloud, maintaining weak positive dependence that goes from $\machav \approx 1$ in parallel filaments to trans-alfvénic regimes in the most perpendicular ones ($\machav\approx2-3$). 
Then, after being impacted by the \hii region the local Mach number increases to values of 2--4 in both simulations, without showing any particular dependence on $\theta_{B\varphi}$. 

This evolution of the Alfvén Mach number highlights how the magnetisation level influences the gas dynamics around filaments, with the weakening dependence on $\theta_{B\varphi}$ in \simBB\ suggesting a transition to a more turbulent regime as the \hii region reshapes the structure of the cloud.

\begin{figure*}[t]
    \centering
    \includegraphics[width=1\linewidth]{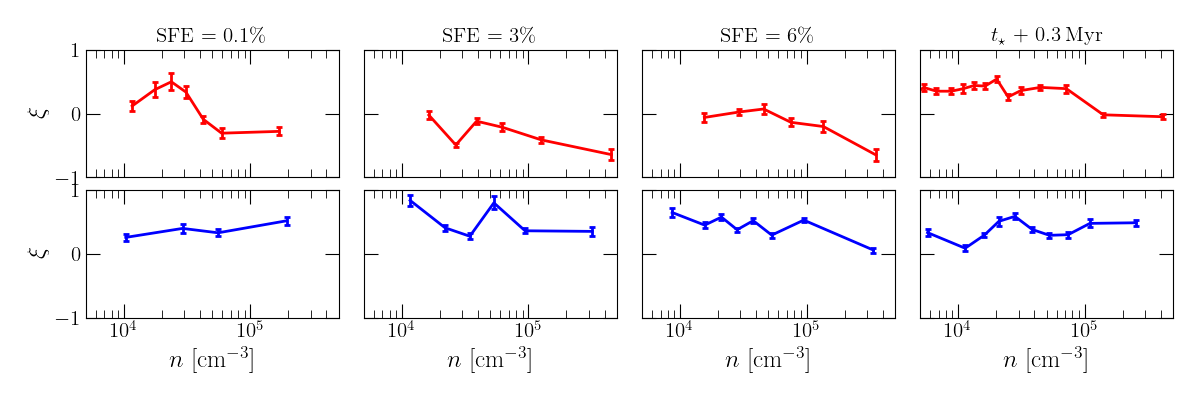}
    \caption{
    Histogram of relative orientation (\citealt{Soler2013}) throughout the simulations \simBB\ (top) and \simA\ (bottom). The graphs highlight the transition from parallel alignments in the low-density regime to more perpendicular ones as the density increases driven by the strength of $B$, which sets the threshold of this transition. From the last column we can evince how the expansion of the \hii region promotes parallel alignments between the density structures and the magnetic field.
\label{fig: HRO}}
\end{figure*}
Further support for the \citet{Soler2017} model comes from examining how the filament density correlates with the relative orientation to the magnetic field. The model predicts that as gravitational forces become more dominant in denser regions, filaments tend to become more perpendicular to $B$, leading to accretion along the magnetic field lines  \citep{Inoue2018}. This behaviour has also been observed in several star-forming regions \citep[e.g.][]{plankcoll2016, Fissel2019}.
To explore this, we exploit the histogram of relative orientation, presented in \citet{Soler2013}. This provides an easy measure to quantify how much the distribution of $\theta_{B\varphi}$ privileges parallel or perpendicular orientation by defining the variable 
\begin{equation}
\xi=P(\theta_{B\varphi} < \theta_\mathrm{crit} ) - P(\theta_{B\varphi} > \theta_\mathrm{crit} ),
\end{equation}
where  $\theta_\mathrm{crit}$ is a reference angle. In observations, this is set to $45^\circ$ as it is applied to 2D angle distribution. To fit it to our tri-dimensional case, we chose $\theta_\mathrm{crit} = 60^\circ$, as this is the average angle obtained from a random distribution. Therefore, negative values of $\xi$ mark distributions are more weighted towards perpendicularity, while positive values indicate a preference for parallelism. We divided our sample of filaments' crest into density bins. For each, we computed the $\xi$ parameter, so as to extract the correlation between the filament density and the relative orientation to the magnetic field.

We present the results in Fig~\ref{fig: HRO}. In \simBB (top row), we can see denser regions become increasingly perpendicular to $B$ before the onset of the \hii region. The sign change in $\xi$ occurs around $n\approx 5\times10^4\,\ndens$, consistent with the predictions of previous numerical studies \citep{Soler2013, Chen2016}. This phenomenon is also observed in real star-forming filaments. While \citet{Fissel2019} provides an estimate for the transition volume density of the order of $10^3\,\ndens$, but with large uncertainties (of the order of a factor of 10 according to the authors), other works retain the transition in column density. \citet{plankcoll2016} provides a column density threshold of the order of $N_{H_2}\approx8\times10^{21}\,\ncoldens$. Assuming a typical width of $0.1\,\pc$ to obtain a rough comparison, we get that our \simBB filaments show a transition at $N_{H_2}\approx1.3\times10^{22}\,\ncoldens$, which is just a factor of 1.5 above from the observed value. 

Concerning the weakly magnetised cloud, $\xi$ shows a slight negative dependence on the density at later times (columns at $\SFE=3\%$ and $6\%$). In this case, a transition is visible at around $5\times10^5\,\ndens$. Again, this well compares with the evolution observed by \citet{Soler2013} in the $\beta=1$ case, where indeed the observed transition shifts towards higher densities by roughly an order of magnitude with respect to their $\beta=0.1$ simulation. 

Finally, after the onset of the \hii region the properties of the filaments change substantially. Overall, the parameter $\xi$ stays positive at all densities, due to the parallel $B$--$\varphi$ alignment observed in Fig.~\ref{fig: angle Bphi SFEHist}. The decreasing trend observed at previous times is preserved only in \simBB, and even in this case, it is much less accentuated. 

\begin{figure*}[h]
    \centering
    \includegraphics[width=1\linewidth]{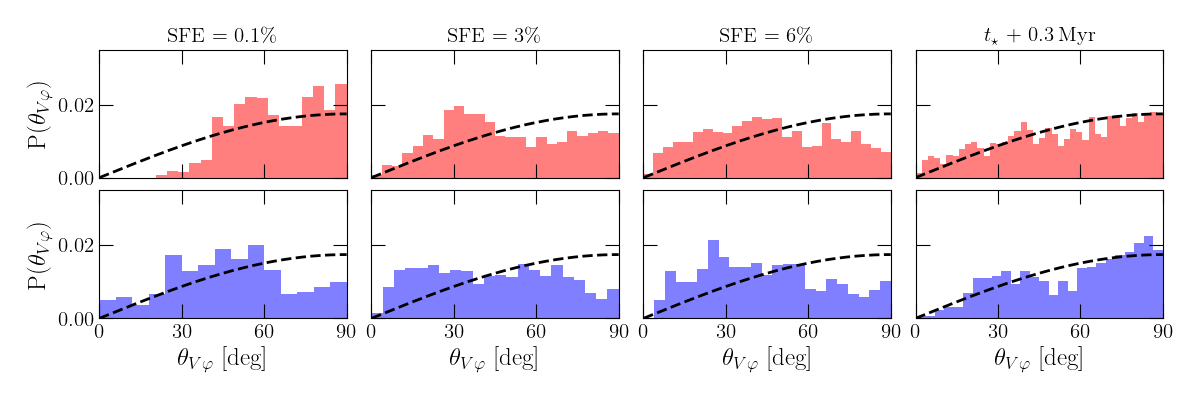}
    \caption{
    Evolution of the distribution of the relative angle between the velocity vector of the filament crest and the filament direction ($\theta_{V\varphi}$) in \simBB\ (top) and \simA\ (bottom). The black dashed lines mark the random distribution for the relative orientation of two independent vectors in 3D. The distributions of $\theta_{V\varphi}$ show a marked tendency of the filaments' crest velocity to align with the filament over time. At earliest stages, the velocity distribution of $\mu=2$ shows a preference towards perpendicular configurations (top left). After the onset of the \hii region, both distributions shift to resemble the random distribution (last column).
\label{fig: angle Vphi SFEHist}}
\end{figure*}
\subsubsection{Filament orientation relative to the velocity}

After examining the influence of the magnetic field on filament orientation, we now turn our attention to the relationship between the velocity field and the filaments, which adds another layer to understanding the dynamics at play. 
Figure~\ref{fig: angle Vphi SFEHist} shows the same evolution presented in Fig.~\ref{fig: angle Bphi SFEHist}, but for the angle between the crest velocity in the lab frame and the filament itself, $\theta_{V\varphi}$. At the onset of SF (first column), the $\theta_{V\varphi}$ distribution varies quite significantly between the two clouds, with \simBB filaments (top) being perpendicular to the flow, and \simA filaments showing a preference for alignment. However, in later times, as SF proceeds, the velocity aligns with the filaments in both simulations. Finally, the onset of the \hii significantly alters the distribution in both simulations, with the two settling on a random-like distribution. This probably suggests that at this stage the expansion of the \hii region dominates the dynamics.

In the weakly magnetised case, the scenario is easily explained with the presence of the hub. As the turbulent support dissipates, the cloud starts assembling the structures and collapsing towards the simulation box's centre. The whole cloud collapses towards the centre, and filaments organise radially around the potential well (Sect.~\ref{sec: cloud structure}). Flowing towards the centre, they accrete from the surroundings and simultaneously feed the central cluster, explaining the preference for parallel alignment. A similar explanation can also be given for \simBB cloud, although the scenario is more complicated. The strong field provides additional support against gravity, and filaments now converge in the multiple intersection hubs spread in the domain. Therefore, on top of the global collapse, the gas flows towards these centres of accretion, affecting the resulting $\theta_{V\varphi}$ distribution. However, we note that in the first phases $\theta_{V\varphi}$ is mainly perpendicular to the density structures, suggesting the presence of strong shocks at the origin of the filamentary network.

After the onset of the \hii region, the distribution suddenly changes in both simulations (last column). Expanding, the bubble collects the density structure on the shell, and the distributions start to lean towards high $\theta_{V\varphi}$, as one could expect in a situation where filaments are compressed on the expanding shell, as seen with observations \citep{Peretto2012, Zavagno2020, Arzoumanian2022}. The distribution now resembles much more the random one, but an analysis performed at later times shows that this is only temporary, since $\theta_{V\varphi}$ continue to shift towards more perpendicular configurations as the \hii region continues expanding. This scenario supports the idea that the alignment of $\theta_{B\varphi}$ found in the previous section is indeed caused by the compression of the field lines along the shock front.

\subsection{Filament dynamics}
\begin{figure}[t]
    \centering
    \includegraphics[width=1\linewidth]{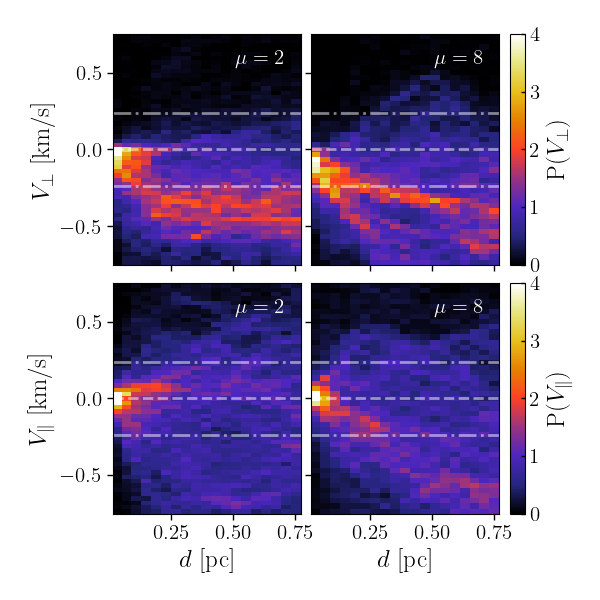}
    \caption{\label{fig: colormap_offset_velocity}
    Distributions of the density-weighted average of the perpendicular (top row) and parallel (bottom row) component of the surrounding gas velocity relative to the filaments' one at varying offsets from the filament crest. The dot-dashed lines give a reference sound speed of $0.24\,\kms$, proper of a molecular medium at 10\,K. The two simulations \simBB (left column) and \simA (right column) are shown at $\SFE=3\%$. From the heatmaps, it is visible how the bulk of $\vperp$ distributions (top) drop to negative at larger offsets, that is infalling,  velocities at larger offsets, stalling around $-0.5\,\kms$. In $\vpar$ distributions, a similar trend is visible in $\mu=8$ filaments, where the negative velocities individuate faster flows inside of the filament than outside. Such behaviour is less visible in $\mu=2$ filaments.
}
\end{figure}

Differences between the \simBB and \simA simulations are also present when examining the dynamical properties of the gas around the filaments. Indeed, the way the gas behaves in the vicinity of the crests is a useful tracer for understanding the underlying physical processes that govern the filaments' evolution and growth. Indeed, we have already demonstrated above how the magnetic field, the velocity field, and the ionising radiation intertwine to determine the evolution of the density structures. Their interplay does not only reflect on the crest of the filament -- the densest and innermost part -- but also on the surrounding environment.

Thus, in this Section, we focus on the gas's local motion surrounding the filaments. For each crest-cell belonging to a filament, we extracted cylindrical shells with a width of approximately $0.04\,\pc$, as described in Sect.~\ref{sec: methods_fil_extraction}.  Within these shells, we characterise the relative motion (subtracting the filament's crest velocity, $V^\mathrm{c}$, to the local gas velocity) in the perpendicular and parallel velocity components with respect to the local filament orientation. Then, we derive the density-weighted average of $\vperp$ and $\vpar$ within each cylindrical shell as a function of the radial distance ($d$) from the filament crest. We discard cells having magnitudes of velocities greater than $5\,\kms$ that could be affected by the presence of jets. As a convention, we take $\vperp$ to be positive when pointing away from the filament and negative when pointing towards it. Therefore, a negative $\vperp$ indicates an infall of the gas onto the filament, either driven by external compression or accretion processes. For $\vpar$, the choice of the sign is more arbitrary, as there is no natural preferred direction along the filament. 
To address this, we compare $\vpar$ with the filament crest velocity. Projecting $V^\mathrm{c}$ onto the filament direction $\phi$, we defined the relative velocity $\vpar$ positive when parallel to $V^\mathrm{c}_\mathrm{\parallel}$, and negative when antiparallel to it.
In other words, the sign is positive if the surrounding gas moves in the same direction as the filament flow but at a higher speed.

Fig.~\ref{fig: colormap_offset_velocity} illustrates these distributions at $\SFE=3\%$ in the two simulations ($\vperp$ in the top panels, $\vpar$ in the bottom panels; \simBB in the first column, \simA in the second). Each column of pixels in the heatmaps displays the normalised distribution of the given velocity component at the corresponding offset from the filament crest. Therefore, the graphs allow us to see how the centre of the distributions (the brightest pixel in the column) varies as we move away from the centre of the filament. The dot-dashed lines provide a reference for the sound speed in the medium ($\approx 0.24\,\kms$ at 10\,K).

Even without considering the temporal evolution -- for now --, some interesting trends are immediately apparent. In both simulations, $\vperp$ consistently peaks at negative velocities, indicating that the filaments accrete gas from their surroundings. When comparing the two simulations, we see that in \simBB, the bulk of the infalling gas approaches the transonic regime further from the filament -- roughly $0.2\,\pc$, compared to $\approx 0.1\,\pc$ for \simA\ filaments. This suggests that strong magnetic fields somewhat inhibit the accretion onto the filaments.  We note that the radial velocities reported in the pure-hydro simulations by \citet{Smith2016} ($-0.5\,\kms,\leq \vperp\leq 0$) are consistent with our findings. Similar magnitudes for line-of-sight velocity dispersions have also been observed in star-forming filaments \citep{Hacar2016, Trevino2019, Chung2021}.
Farther out, the velocity distribution remains close to the sound speed, but the spread increases as the gas decouples from the filament's influence. Closer to the filament crests, the accretion proceeds subsonically, suggesting that accretion is more gradual near the centre. 

For what concerns $\vpar$ instead, the distributions appear more chaotic. Overall, the distributions tend to have more extended tails towards negative values, meaning that the gas within the filament tends to flow faster than the gas outside. While in \simBB, the bulk of the distribution (the brightest pixels) remains around zero, in \simA, we can see a clear trend towards negative $\vpar$. However, as we get farther from the crest, the gas loses correlation with the analysed filaments, and the tails spread.

\begin{figure*}
    \centering
    \includegraphics[width=0.9\linewidth]{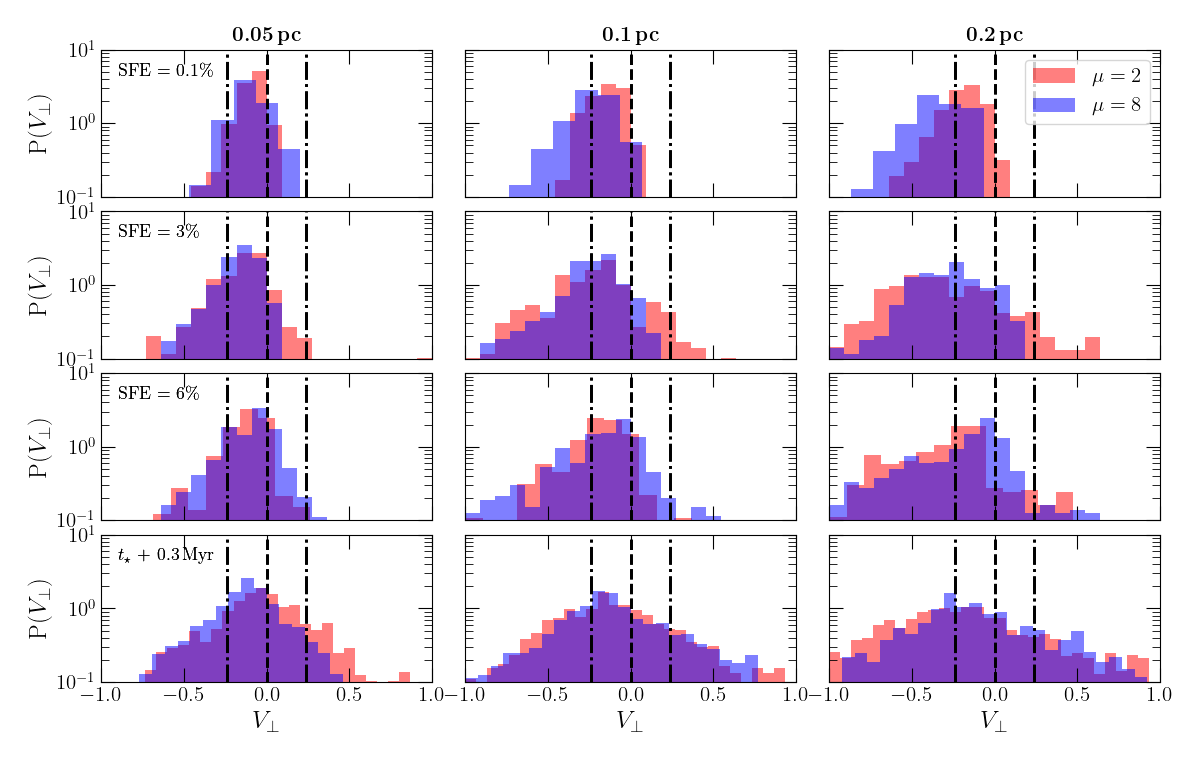}
    \caption{
    Evolution of the distribution of the perpendicular velocity $\vperp$ of the surrounding gas relative to the local filament frame in the two simulations \simBB\ (red), \simA\ (blue). Looking across the columns, the distributions are shown at three different radial offsets from the filament crest ($0.05\,\pc,\, 0.1\,\pc,\, 0.2\,\pc$), while the temporal evolution is displayed across the rows at 
    different $\SFE$ of the cloud ($0.1\%,\, 3\%, 6\%$) and $0.3\,\Myr$ after the onset of the main \hii region (last row). The black dot-dashed lines give a reference sound speed for a molecular medium at $10\,$K. The overall preference towards negative $\vperp$ highlights how filaments never stop accreting. The widening of the wings during the feedback-dominated phase suggests more random velocities (cf. Fig.~\ref{fig: angle Vphi SFEHist}).
\label{fig: perp velocity hist}}
\end{figure*}

\begin{figure*}
    \centering
    \includegraphics[width=0.9\linewidth]{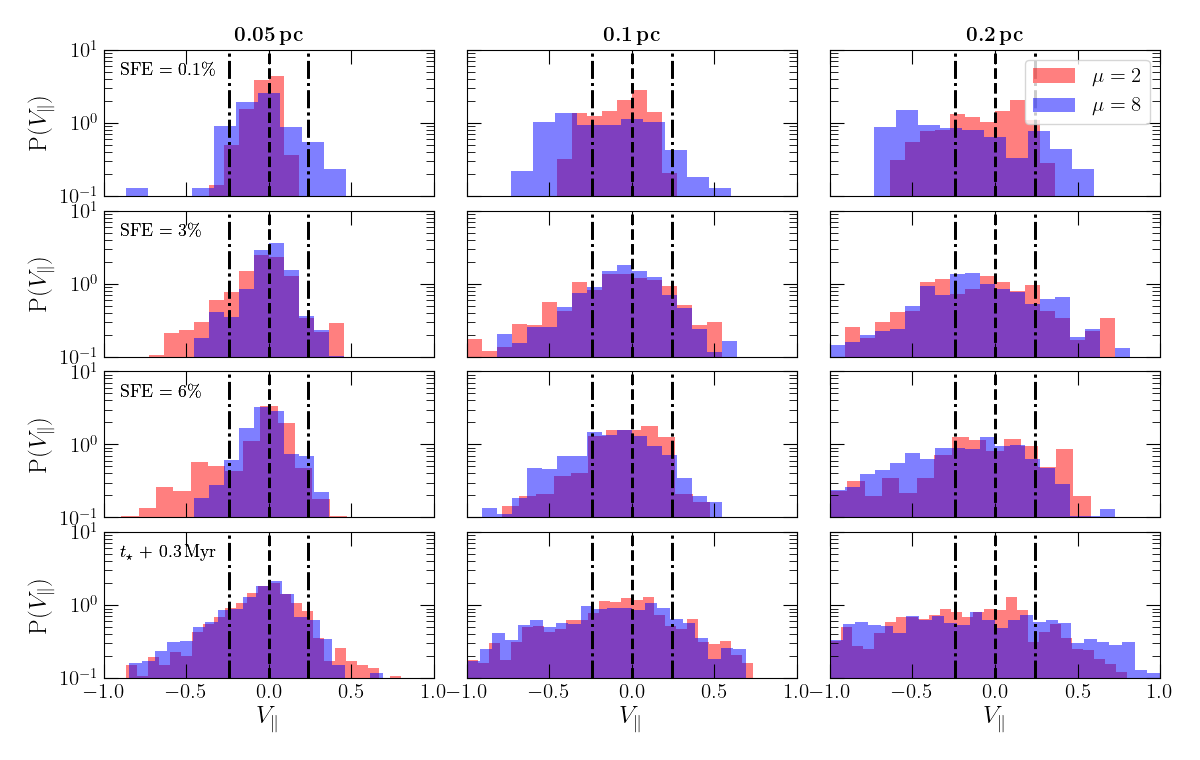}
    \caption{
    Same as Fig.~\ref{fig: perp velocity hist} but for $\vpar$. The sign of $\vpar$ is decided according to the verse in the lab frame of the filament crest, projected on the filament direction $\varphi$. The extended negative wings indicate that the gas within the filament exhibit faster flows than the background.
\label{fig: para velocity hist}}
\end{figure*}

To gain a more detailed understanding of these trends, we focus on the temporal evolution of three specific radial distances from the filament crests -- at $0.05\,\pc$, $0.1\,\pc$, and $0.2\,\pc$ -- which represent the inner, edge, and outer regions of the filament environment. Figs.~\ref{fig: perp velocity hist} and \ref{fig: para velocity hist} show how these velocity distributions evolve throughout the simulations at the same time steps as in Fig.~\ref{fig: angle Bphi SFEHist}.


\subsection{Perpendicular motion}
Looking at the distributions at $0.05\,\pc$ (first column of Fig.~\ref{fig: perp velocity hist}), we observe minimal variations between the two simulations. The velocities are mostly subsonic, with slight variation in the position of the peak, although filaments in the weakly magnetised simulation show a slight shift towards more negative velocities. This suggests that the filaments are undergoing a slow collapse. We note that the wings of the distributions are broader in the $\mu=8$ case, especially in the first phases, indicating a more turbulent environment. As expected, once the bubble expands and sweeps up the gas the wings extend in both cases. 

The similarity of the distributions indicates that the magnetic field plays a secondary role in determining the internal dynamics of the filament at these offsets. This conclusion is supported if we look at Fig.~\ref{fig: angle Bphi SFEHist}, showing the Alfvén Mach number near the filament crests.
In \simA, $\machav$ remains consistently above unity across the filament distribution, indicating super-alfvénic turbulence. In \simBB, $\machav$ quickly increases to trans-alfvénic regimes as the cloud evolves, suggesting that the magnetic field's influence at small scales is most significant at earlier times. 
Afterwards, the significant changes in the distributions following the passage of the \hii region confirm that it is the ionising flux, rather than the magnetic field, that governs the local dynamics at this stage. 

Further out from the crest (second and third columns), the distributions of the two simulations display more noticeable differences.  At the onset of the SF event, $\vperp$ shifts towards more negative velocities in \simA, where the weaker magnetic field allows stronger compressions. The shift is instead limited in \simBB, where the stronger $B$ inhibits the converging flows. However, as the cloud evolves, the peak of the $\vperp$ distribution in weakly magnetised filaments moves to subsonic intensities, possibly translating into a reduction of the accretion rate of such structures. Conversely, the distribution of \simBB shifts towards the negative velocities, and by $\SFE=6\%$ the magnitude of this motion becomes greater than in \simA filaments at $0.1\,\pc$. This is compatible with the results of Fig.~\ref{fig: densities} presenting the filament density distributions. Initially, the magnetic support suppresses the accretion onto filaments, and this results in less dense filaments compared to those in \simA (first panel). As gravity overcomes the magnetic pressure, the density distribution of the two clouds resembles each other. The delay between the two simulations remains relatively fixed to $0.3$--$0.4\,\Myr$ throughout the simulation, and this means that filaments in the strongly magnetised cloud have to achieve higher accretion rates in order to overcome the initial phase of quiescence.

These findings, combined with the behaviour observed at $0.05\,\pc$, prove that the intensity of the magnetic field plays a crucial role during the early phases of the filaments' evolution, slowing down their evolution and growth. However, as time progresses, gravity begins to dominate, and accretion is favoured along the magnetic field lines. At $0.2\,\pc$, we observe supersonic negative velocities, which gradually become subsonic as they approach the filament crest, which is a clue for rapid accretion modes as suggested by observations \citep{Lu2018, Chung2021}.  Additionally, we note that the skew of the distributions increases as the SF goes on,  with the negative wing growing more prominent, further indicating that the filaments never stop accreting from the surrounding gas.

After the eruption of the \hii region, we observe a sudden change in behaviour (last row). The compression generated by the expanding ionised bubble causes a spreading in the wings of the distributions. Comparing the last column of Fig.~\ref{fig: angle Vphi SFEHist}, we can imagine the origin of this behaviour to be the more energetic and chaotic environment stimulated by the expansion of the shell. However, we can still register a preference towards negative $\vperp$, which is consistent with the presence of compressive flows driven by the shell. 

Finally, comparing distributions in \simA at all offset with the ones at $\SFE=6\%$, we also see a shift towards negative velocities of the distributions, highlighting the compression applied by the ionising radiation. In \simBB the peaks at the earlier snapshot are already negative ($\approx -c_\mathrm{s}$ at the largest offsets), and the enhanced converging flows are not registered. However, in the innermost part, we note a peak position compatible with zero, possibly caused by the additional support against compression by $B$.

\subsection{Parallel motion}
\begin{figure}
    \centering
    \includegraphics[width=1\linewidth]{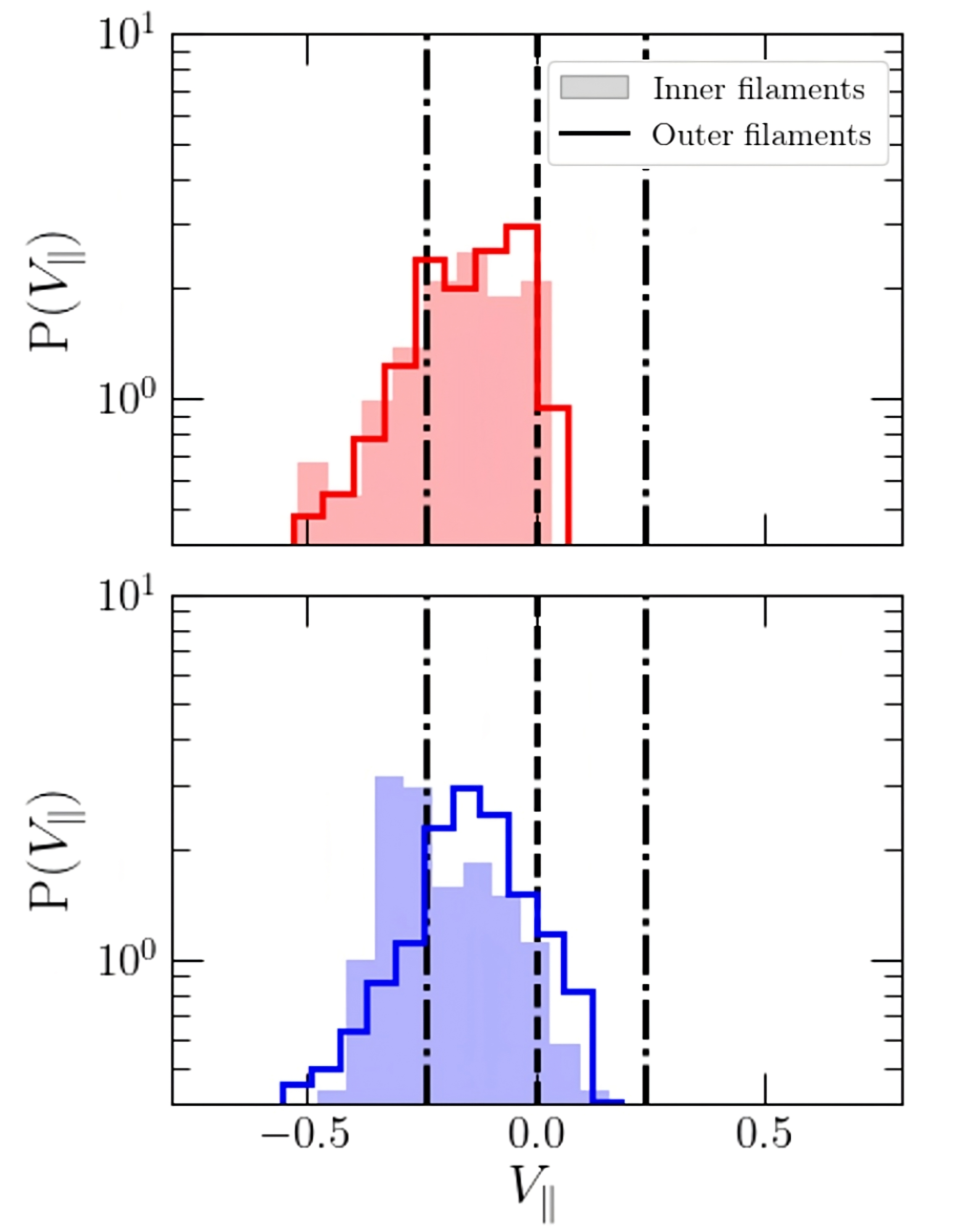}
    \caption{
    Distributions of $\vpar$ in the two simulations (\simBB top, \simA bottom) at an offset of $0.1\,\pc$ from the filaments crest when $\SFE=3\%$. The set has been divided between filament sections closer (filled histogram) and farther (hollow histogram) than $2\,\pc$ from the cloud centre. The black dot-dashed lines give a reference sound speed for a molecular medium at $10\,$K.   All four distributions display a shift towards negative velocities (faster flows along the filaments relative to the surrounding gas). However, the shift in peak velocity between the inner and outer distributions, evident in \simA but absent in \simBB, suggests that in a well-developed HFS the faster flow in the filaments is driven by the central potential well. This supports the conveyor belt scenario, where filaments efficiently channel material toward the hub.   
\label{fig: histVpar with distance}}
\end{figure}

In addition to the perpendicular motion, the motion of the surrounding gas in the direction parallel to the filament provides further insights into the evolution of these structures. In particular, observations of HFSs suggest matter flow along the filaments onto the hubs \citep[e.g.][]{Kirk2013, Peretto2013, Ma2023}. We aim to investigate the presence of such motions by analysing the parallel velocity in our simulations. As shown in Fig.~\ref{fig: para velocity hist}, prior to the birth of the most massive star, the peak of the velocity distribution for $\vpar$ tends to be centred around zero, especially when looking at the $0.05$ and $0.1\,\pc$ columns. This indicates that the overall gas flow is generally uniform in the proximity of the crest. However, we also notice that both the \simBB and \simA distributions are predominantly skewed towards negative velocity, and that this skewness increases as we move out from the centre of the filament. The tendency to negative velocities becomes more evident as the clouds evolve. As for the $0.2\,\pc$ column, negative shifts of the peaks can be observed in \simA at earlier times compared to the distributions from the inner regions, with even more accentuated asymmetry in the distributions. This preference for negative velocities is consistent across the snapshots at different times of both simulations. 

Negative parallel velocities of the gas at $d\gtrsim0.1\,\pc$ indicate that the gas surrounding the filaments flows slower than the gas within the filament. The imbalance between the negative and positive wings supports the idea that filaments act as conveyor belts, channelling gas efficiently towards local gravitational wells, as proposed by \citet{Longmore2014} and \citet{Krumholz2020}. Hints of such behaviour have also been observed in real clouds through the study of the line of sight velocity gradients \citep{Zhang2023, Rawat2024}.

To further investigate this hypothesis, we extracted $\vpar$ distributions at varying distances from the centre of the system, chosen to be the minimum of the potential well as in Sect.~\ref{sec: cloud structure}. As a reference, we chose the snapshots at $\SFE=3\%$, where the HFS has already developed in \simA, but has yet to form in \simBB. Despite the similarity of the distributions at this stage (second row of Fig.~\ref{fig: para velocity hist}), if the hypothesis is correct, the velocity difference should be more pronounced closer to the central cluster in the low-magnetised cloud, as the filament flow will be directed radially (Fig.~\ref{fig: radial angle}), and towards the minimum of the potential. On the contrary, in \simBB, the filamentary cloud presents multiple star-forming hubs. If the flow along the filament is directed towards these accumulation points, the location of the potential minimum does not play a relevant role in determining the motion along the filaments, and we do not expect significant differences between the dynamics of filaments close and far from the centre. 

Fig.~\ref{fig: histVpar with distance} illustrates the distributions for the two simulations at $0.1\,\pc$ offset (corresponding roughly to the edge of the filaments), distinguishing between filament portions closer (filled histogram) and farther (hollow histogram) than $2\,\pc$ from the minimum of the potential well. The threshold was chosen to bring roughly comparable sample sizes for both distributions in both simulations. Little differences can be found in \simBB, while in \simA filaments closer to the centre exhibit faster motion relative to the background compared to those further out, supporting the idea that the hub-filament structure plays a crucial role in regulating gas flow and filament evolution. 

To have a quantitative confirmation, we perform a Mann–Whitney U test\footnote{The Mann-Whitney U test is more suitable here than the KS test used above, as it focuses on differences in central tendency rather than on the overall distribution shape. This allows to accurately determine whether the observed shifts of the parallel velocities are significant or not.} to assess whether the overall velocities closer to the central cluster differ significantly from those farther out, with the alternative hypothesis that the relative velocity of the flow around filaments is more negative when those are closer to the cloud's centre.

In the strongly magnetised case, the test returns a $p$-value of 0.3, confirming that the velocity distributions show no significant differences between the inner and outer regions. As the SF centres are spread over the domain (several clusters can be noted in the second panel of the first row in Fig.~\ref{fig: cloud_simulation}), the centre of the cloud does not play any particular role during the earliest stages of the simulation. In contrast, in \simA, the test produced a $p$-value well below the significance threshold ($p\approx 10^{-5}$), confirming that filaments near the central hub exhibit significantly higher velocities than those farther out.

\section{Discussion} \label{discussion}
Our study reveals how the complex interplay of the magnetic field strength, the large-scale cloud configuration, and the feedback from massive stars heavily influence the evolution of filamentary structures in star-forming clouds. 
In this Section, we summarise our results highlighting the role played by these three elements. First, we discuss how the strength of the magnetic field drives distinct evolutionary pathways both concerning the overall cloud morphology and the single filaments (Sect.~\ref{subsec: impact mag}). Next, we consider how the cloud morphology, tightly linked to the initial cloud magnetisation, regulates the gas flow and shapes the filament dynamics (Sect.~\ref{subsec: role morpho}). Then, we examine how the expanding \hii region affects the environment, influencing the filament structures and altering the overall gas dynamics (Sect.~\ref{subsec: impact rad}). Finally, we discuss the possible limitations in Sect.~\ref{subsec: limitations}.

\subsection{Impact of magnetic field strength}
\label{subsec: impact mag}
The strength of the magnetic field plays a decisive role in shaping the global structure and dynamics of our simulated star-forming clouds. In \simBB, the strong magnetic field resists gravitational collapse, leading to the formation of multiple hubs, which only at later times merge into a single HFS. In the low-magnetisation case, instead, a single central hub is quickly formed with filaments radially converging towards the centre of the cloud (Fig.~\ref{fig: radial angle}). 

In our simulations, two distinct modes of filament alignment emerge depending on the strength of the magnetic field (Sect. \ref{sec: orientation of filament}, Fig.~\ref{fig: angle Bphi SFEHist}). In the strongly magnetised case, filaments tend to align perpendicular to the magnetic field lines, while in the weakly magnetised case, filaments are generally aligned with the magnetic field. This is because the field lines in the stronger field resist being bent by the turbulent motions of the gas. Thus, the gas flows along the field lines, and leads to the formation of structures perpendicular to them. In weak magnetic fields, instead, turbulence and gravity can drag the magnetic field line \citep{Gomez2018, Inoue2018, Pineda2023}, promoting a parallel alignment with the magnetic field. This is consistent with theoretical expectations from the simple model proposed by \citet{Soler2017}, and well agrees with the picture described in the larger-scale study by \citet{Seifried2020}. 

These two configurations persist throughout the evolution of the cloud. In \simBB, the filaments remain mainly perpendicular to the magnetic field prior to the onset of the \hii region, while in \simA, the parallel configuration is preferred, suggesting that, once established, these configurations are stable \citep{Soler2017}. 
This may offer a way of classifying filaments observed in star-forming regions, since it appears that whether a star-forming filament is aligned parallel or perpendicular to the magnetic field heavily depends on the initial conditions of the system\footnote{We highlight that here we are discussing the densest filaments of the region, that is the ones that took part to the SF process \citep{Fissel2019}. In strong magnetic field environments, striations parallel to the magnetic field are usually observed \citep{Palmeirim2013}. However, those structures are usually less dense and starless. Therefore, they are not retrieved with the \disperse parameters adopted in Sect.~\ref{sec: methods_fil_extraction}.}. 

Additionally, we found that filaments that are more perpendicular to the field tend to be denser (Fig.~\ref{fig: HRO}). This  correlation is more evident in \simBB, hinting at an active role of the magnetic field in determining the accretion process (a parallel alignment suppresses perpendicular accretion from all directions). 

The connection with the accretion process is also confirmed by comparing the distributions of $\vperp$ in the two simulations (Fig.~\ref{fig: perp velocity hist}). This also relates to the observed Alfv\'en Mach number in the interiors of the filaments. We found that initially the filaments of \simBB display a narrower distribution of $\vperp$ at all offsets compared to the \simA filaments. Simultaneously, we see from the dependence of $\machav$ from $\theta_{B\varphi}$ (yellow line in Fig.~\ref{fig: angle Bphi SFEHist}) that the magnetic field play a most dominant role in the earliest stages of SF, as the Alfv\'en Mach number reaches values as low as $0.2$ in the most parallel filaments (see also Appendix~\ref{sec: appendix before SF}). This, connected with the dependence of the density from $\theta_{B_\varphi}$ (Fig.~\ref{fig: HRO}), demonstrates a specific evolutionary path that further favours perpendicularly aligned filaments. Such complicated intertwining of accretion rate, density, magnetic field strength and relative orientation is, if not absent, less critical in \simA, as the $\machav$ always sets on values above one, without any particular temporal evolution of the relation with $\theta_{B\varphi}$.

Only at later stages does the distribution of $\vperp$ spread in \simBB filaments and start to resemble that observed in \simA ($d=0.05\,\pc$, $\SFE>3\%$). The similarity in the two distributions can also be observed in Fig.~\ref{fig: angle Bphi SFEHist}, showing the inner part of the filament settling on a mildly trans-alfv\'enic regime in the filaments in both simulations. However, we can tell that $B$ still has an active role in determining the external dynamics, as at outer offsets, the wings spread wider, and the bulk moves to more negative velocities compared to the distribution from weakly magnetised filaments, suggesting higher accretion rates. 

\subsection{Consequences of cloud morphology}
\label{subsec: role morpho}
In Sect.~\ref{res} we showed that the also overall morphology of the cloud impacts the evolution of filaments, especially in determining the flow dynamics and the subsequent SF processes. The development of a single, central HFS, as observed in \simA and later in \simBB, introduces a distinct large-scale structure that alters how the gas flows.
As the central hub forms, filaments merge and dispose radially around the central well (Fig.~\ref{fig: radial angle}). This structured flow is reflected in the gas dynamics, with material moving preferentially along the filaments towards the hub, thus promoting a parallel alignment between filaments and the gas velocity (Fig~\ref{fig: angle Vphi SFEHist}). This goes along with the several observations showing a preferred alignment of the magnetic field with the filaments in HFSs -- as in Monoceros R2 \citep{Hwang2022}, the Cocoon Nebula \citep{Wang2020}, the hub of NGC 6334 \citep{Arzoumanian2021} and the Serpens Core \citep{Sugitani2010, Pillai2020}.

Moreover, we find that filaments generally exhibit faster motion relative to the background, further supporting the idea that these structures act as efficient channels for mass transport. This is noticeable in the $\mu=8$ simulation, where the central hub provides a reference system (Fig.~\ref{fig: histVpar with distance}). Indeed, the filament--background velocity difference increases towards the centre, suggesting that the hub not only organises the spatial structure of the cloud but also regulates the flow dynamics at larger scales.

On the other hand, in \simBB, the absence of a dominant central hub results in a more distributed network of filaments, where SF proceeds simultaneously in multiple centres. This more dispersed configuration leads to a slower, more gradual collapse. Instead of radially converging towards a central cluster, as in \simA, the filaments exhibit less organised gas flows on a large scale since the multiple hubs direct the accretion flows towards separated accumulation points. The scenario varies in the later stages, when turbulence dissipates and gravity takes over the magnetic support. Towards the end of the simulations, the various clusters merge, eventually resembling the HFS observed in \simA. This supports the hypothesis that the two stages -- filamentary cloud and HFS -- are, in fact, different evolutionary stages of clouds' life \citep{Schneider2012, Li2014}. 

\subsection{Effects of ionising radiation}
\label{subsec: impact rad}
The onset of the \hii region marks a turning point in the evolution of the cloud. As soon as the ionising radiation impacts the filaments, it drastically alters their properties. Not only does the alignment with the magnetic field shift, but the Alfv\'en Mach number within the filaments and the surrounding dynamics are also affected.

As the bubble expands, the front advances at the sound speed of the ionising medium ($\approx 10\,\kms$), and compresses the magnetic field lines onto the density structures, promoting a parallel configuration. While this does not significantly enhance the already-present parallelism of \simA filaments, we report a sudden transition from a predominantly perpendicular orientation to a parallel one in the strongly magnetised case. This scenario is compatible with the observed $B$ field aligned with the dense shell around \hii regions \citep{Arzoumanian2021, Tahani2023}.
In our simulations, we can observe the magnetic field aligning with both the structures, stretching along the edges of the shock front and the pillar-like formation extending towards the central region  \citep[Fig.~\ref{fig: pillars}; ][]{Pattle2018}.

The ionising radiation also visibly affects the gas dynamics within and around the filaments. As the \hii region expands, the velocity dispersion and the local $\machav$ within the filaments significantly increase. This is evident in Fig.~\ref{fig: perp velocity hist}, where the passage of the shock front leads to an increase in infall velocities, caused by the compression of the gas rather than an actual accretion process as seen at previous times. 

The \hii region causes the SF event to shut down in both clouds. In \simBB, the sparse morphology allows the ionising bubble to expand faster, and after only $0.5\,\Myr$, the SF rate results are quenched by a factor of 10. In \simA, instead, the dispersion process is slightly slower, so that by $0.5\,\Myr$ the SF rate is reduced to $\approx 200\,\MSun\, \Myr^{-1}$ (quenched by a factor of 5). In any case, by the end of the SF event, both the simulations end up with similar SF efficiency (12 and 13\% after $\approx 1\,\Myr$ from the onset of the \hii region). 

\subsection{Limitations}
\label{subsec: limitations}
Numerical simulations are powerful tools for exploring the temporal evolution of filamentary structures in star-forming molecular clouds. They provide invaluable insights into the dynamic processes that govern filament formation, growth, and interaction with their environments, which are challenging to capture through observations alone. However, those are not free from limitations. For example, the idealised setup we need to start the simulation represents a double-edged sword. On the one hand, it allows for a complete control on the environmental parameters and provides a way of disentangling the impact of a given mechanism by simply changing its value at the start of the simulation (as we did for the magnetic field). On the other hand, however, real clouds are heterogeneous entities, and the physical parameters observed can vary many orders of magnitude that cannot be explored simultaneously with a limited set of simulations  -- we mention \citet{Klassen2017}, \citet{Dhandha2024}, and \citet{Chira2018} for studies covering the impact on the different properties of the cloud mass, the turbulence modes (compressive, solenoidal or mixed) and the surrounding environment.

This also extends to the treatment of massive stars within the cloud. Due to computational constraints, their formation is not self-consistently coupled with the gas evolution. Instead, these stars are introduced in the sink particles at predetermined stages, as the resolution in our simulations does not permit accurate sampling of the high-mass tail of the initial mass function (see Sect.~\ref{sec: stellar feedback}). While the radiation field generated by these stars is reliably modelled, the exact timing, location, and masses of the ionising stars that would naturally form in a real cloud could influence the specific evolutionary path of the cloud.
Nevertheless, our findings remain robust in capturing the overall impact of massive stars and their radiation fields on the global dynamics and evolution of the molecular cloud.

Another imperfection is the absence of a larger environment in which the clouds reside. Real clouds form from and interact with a larger medium, which can take an active part in their evolution. This isolation in our simulations means we cannot account for external factors such as large-scale shear flows \citep{Abe2021} and interactions with neighbouring clouds \citep{Ma2022, Maity2024}.

This limitation, however, leaves space for speculation on the impact of the clouds' evolutionary speed in a more realistic environment. Indeed, we find that the presence of a stronger magnetic field significantly delays the formation of the HFS, intended as the presence of a clear central cluster dominates the overall potential. If the interaction with the surroundings can alter the cloud's morphology, and even disrupt it, this would mean that the formation of an HFS may be favoured in environments where the ambient magnetic field is weak. Vice-versa, since the filamentary structure phase would last longer in a highly magnetised environment, wherever the magnetic field is more intense, we would expect a predominance of clouds in the filamentary stage. 
Further cloud-scale observations of the matter distribution (filamentary or HFSs) and the magnetic field geometry and strength may provide important constraints on the impact of the magnetic field on cloud evolution and its SF and dispersal history.

Recent simulations adopting the zoom-in technique \citep{Dobbs2015, Seifried2017} offer a potential solution to this limitation. Although computationally expensive and more challenging due to the need to incorporate various physical processes across different scales (and avoid possible numerical artefacts, as discussed in \citealt{Seifried2015}), zoom-in simulations can bridge the gap left by current isolated collapsing cloud models, opening the path for future studies on the evolution of the filamentary structure within the star-forming molecular clouds. A relevant example is \citet{Seifried2020}, who applied this approach to examine the orientation of filaments relative to the magnetic field. There, they report similar dependencies of $\theta_\mathrm{B\varphi}$ on density and magnetic field strength, reinforcing the physical reliability of the findings of the present study.

\section{Conclusions} \label{conc}
Through MHD numerical simulations of a star-forming molecular cloud immersed in different magnetic environments, we have investigated how magnetic fields and stellar feedback influence the growth and evolution of filamentary structures arising from a gravoturbulent environment. 
The main results are: 
\begin{itemize}
    \item
    The global structure of the cloud evolves differently under the two magnetisation cases. In \simBB we observe the formation of multiple, separated hubs of SF, while in \simA, the cloud features the early formation of a HFS. The stronger magnetic field supports the cloud against rapid collapse, resulting in a more distributed filamentary network and a slower evolution.
    \item The different strengths of magnetisation lead the cloud to evolve at a different pace, with the run with stronger $B$ displaying a delay of $\approx0.4\,\Myr$ compared to \simA simulation. 
    \item Although the structure of the cloud and the evolution timescale are different, the SFE in the cloud is similar at the end in the two magnetisation cases, of about 12--13\%. 
    \item 
    Looking at the evolution of dense filaments, we observe significant differences in the distribution of the relative angle between the magnetic field and the filament orientation. In the high-magnetisation case, filaments exhibit a more pronounced perpendicular orientation during the early evolution of the cloud, while the low-magnetisation case favours parallel orientation. Over time, these distributions become more pronounced, enhancing their perpendicular or parallel configurations in \simBB\ and \simA, respectively, as predicted by the analytical model of \citet{Soler2017}.
    \item Supporting the analytical model of \citet{Soler2017}, we find a positive correlation between filament density and the relative angle $\theta_{B\varphi}$. Denser structures tend to align more perpendicularly to the local magnetic field, and exhibit higher $\machav$. This is caused by gravity becoming increasingly dominant in denser regions, overcoming magnetic support and influencing filament orientation.
    \item The introduction of radiative feedback from high-mass stars strongly influences the filament orientations. The expansion of \hii\ regions and the associated ionisation fronts promote a more parallel alignment between filaments and magnetic fields, and this is particularly evident in \simBB, where we can observe a radical shift from perpendicular to parallel alignments. This shows how stellar feedback can modify the magnetic configurations, reshaping filamentary structures and their relative orientations and governing the overall dynamics of the medium. 
    \item 
    The filament dynamics reveal apparent differences in the local gas behaviour in both the parallel and perpendicular directions. Filaments constantly accrete material from their surroundings throughout their evolution -- indicated by negative values of the perpendicular velocity component $\vperp$. The magnetic field plays a crucial role in governing how gas flows towards the filaments. In the early stages, a stronger magnetic field inhibits strong infall velocities onto the filament crests. In later stages, however, filaments in the strongly magnetised run experience higher accretion rates.
    \item 
    We observe differences in the surrounding gas's parallel velocity component $\vpar$ relative to the filament motion. Generally, gas inside the filament flows faster than gas outside. A central hub enhances this velocity difference, with filaments closer to the central cluster displaying higher relative velocities than those farther away. This supports the conveyor belt scenario, where filaments act as channels funnelling gas towards star-forming regions.
\end{itemize}

Our study demonstrates that magnetic fields and stellar feedback significantly influence the evolution of molecular clouds, actively shaping the resulting structures and dynamics. Magnetic field strength determines the mode of filaments formation and their orientation relative to the magnetic field, which persists throughout the cloud's evolution until altered by the onset of the main \hii region. The interplay between magnetic fields and gravity affects filaments' density and accretion properties, impacting the SF processes.

\vspace{0.5cm}

\begin{acknowledgements}
We thank the anonymous referee for the precise and constructive comments that helped improve the clarity of the article. PS thanks J.~Soler for his helpful suggestions and stimulating discussions. AZ thanks the support of the Institut Universitaire de France \href{https://www.iufrance.fr/}{(IUF)}. Part of this work was supported by the NAOJ Research Coordination Committee, NINS (NAOJ-RCC-2401-0403).
We thank the High-Performance Computing Division of the San Piero a Grado Green Data Center (University of Pisa), without which this work would not have been possible.
      Software: Aladin \citep{2000A&AS..143...33B}, astropy \citep{astropy:2022}, Matplotlib \citep{matplotlib}, NumPy \citep{numpy}, SciPy \citep{SciPy2020}. This document was prepared using the \href{www.overleaf.com}{Overleaf web application}.
\end{acknowledgements}

\bibliographystyle{aa} 
\bibliography{aa53795-25}

\begin{appendix}
\section{Impact of the persistence threshold}
\label{sec: appendix thr}

\begin{figure}[htb]
    \centering
    \includegraphics[width=1\linewidth]{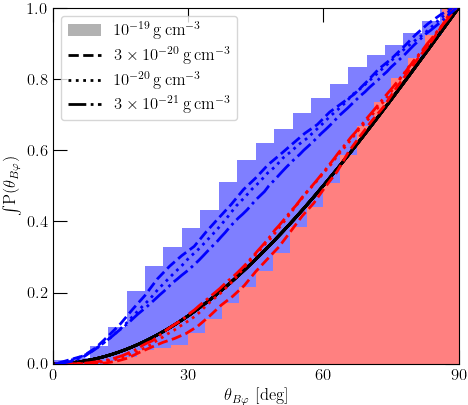}
    \caption{Cumulative distribution function of $\theta_{B\varphi}$ (see Sec.~\ref{res}) for \simBB (red) and \simA (blue), extracted at $\SFE=3\%$ for different persistence thresholds: $10^{-19}\,\dens$ (used in the text, filled histogram), $3\times10^{-20}\,\dens$ (dashed lines), $10^{-20}\,\dens$ (dotted lines) and $3\times10^{-21}\,\dens$  (dash-dotted lines). The black solid line indicates the random distribution.
    \label{fig: app_B} }
\end{figure}

\begin{figure*}[h!]
    \centering
    \includegraphics[width=0.227\linewidth]{Images/Simulations_snaps/mu2/mu2_SFE3.pdf}
    \hfill
    \includegraphics[width=0.227\linewidth]{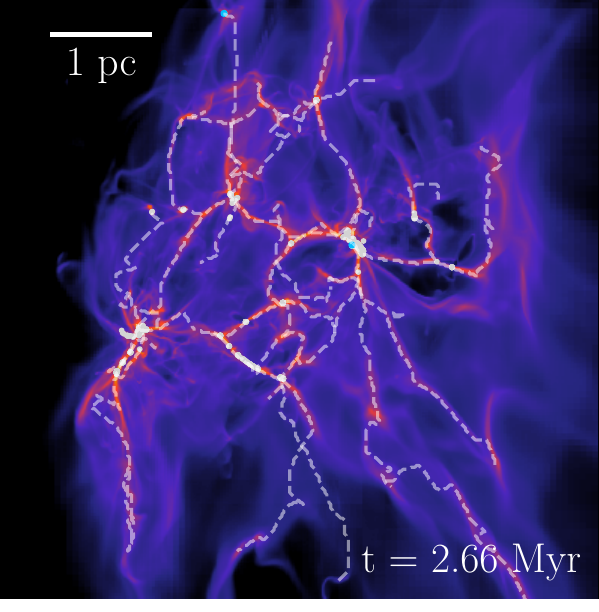}
    \hfill
   \includegraphics[width=0.227\linewidth]{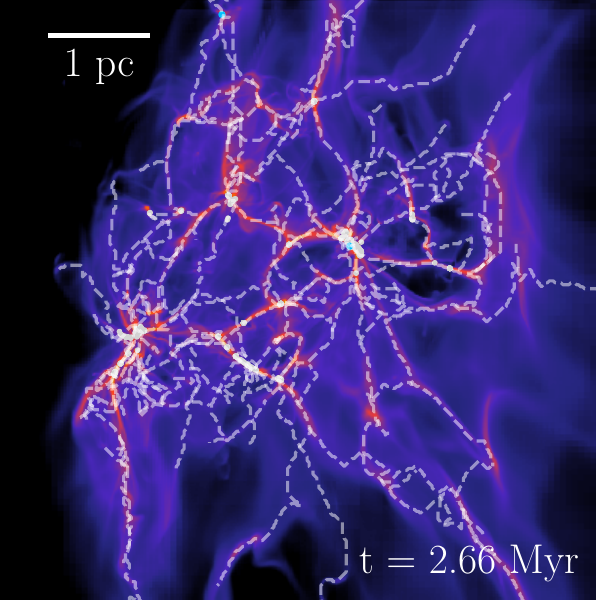}
    \raisebox{-0.02cm}{
    \includegraphics[width=0.29\linewidth]{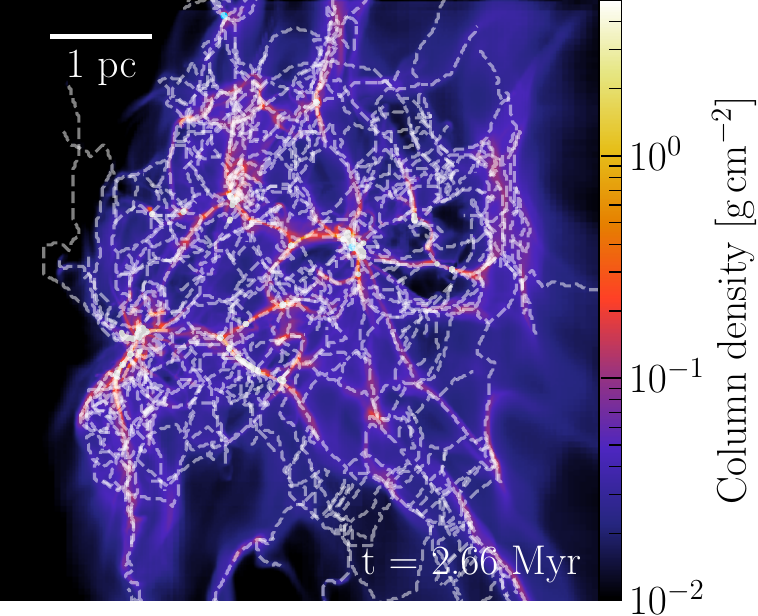}}
    \\
    \includegraphics[width=0.227\linewidth]{Images/Simulations_snaps/mu8/mu8_SFE3.pdf}
     \hfill
   \includegraphics[width=0.227\linewidth]{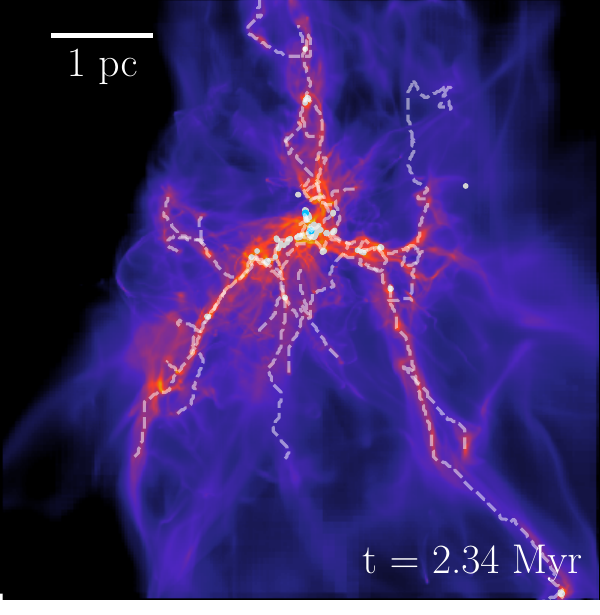}
    \hfill
   \includegraphics[width=0.227\linewidth]{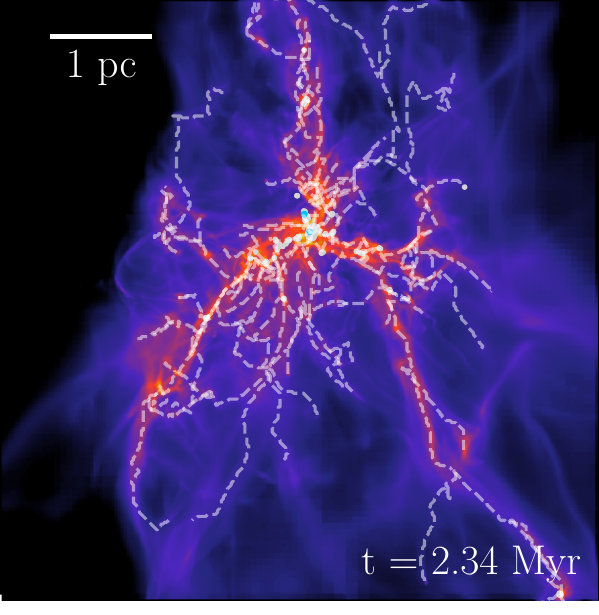}
    \hfill
    \raisebox{-0.02cm}{
    \includegraphics[width=0.29\linewidth]{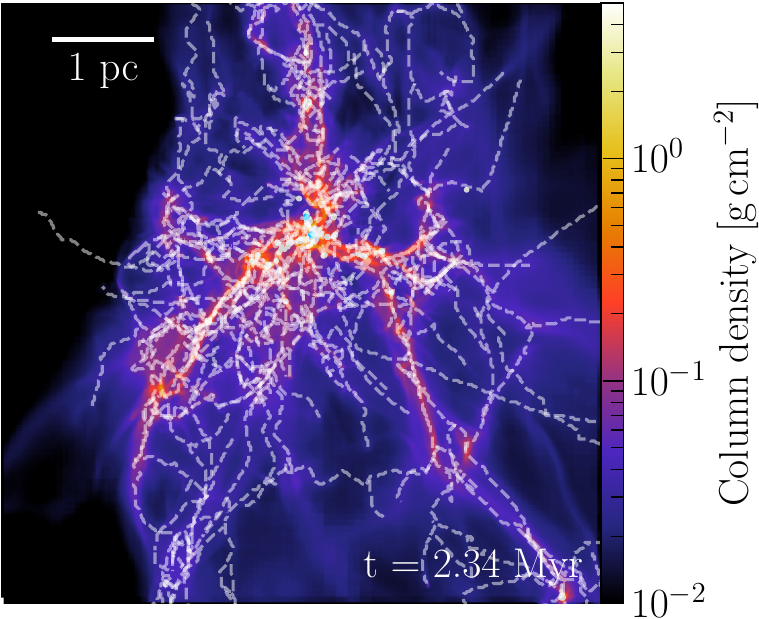}}

    \caption{Progression of the extracted skeleton at various persistence thresholds for the snapshot at $\SFE=3\%$ in \simBB (top) and \simA (bottom). From left to right, the columns correspond to persistence thresholds of $10^{-19}\,\dens$ (used in the text), $3\times10^{-20}\,\dens$, $10^{-20}\,\dens$ and $3\times10^{-21}\,\dens$.    
    \label{fig: filamentary_at_thrs} }
\end{figure*}

To assess the robustness of the result discussed, it is necessary to understand the impact of the code parameters on the filaments. A high persistence threshold such as the one adopted for the analysis assures that the structures found are real features but risks cutting out a consistent fraction of the filaments. For this analysis, we reduced the \disperse input threshold. We explored the impact of the persistence threshold we compared four different values for the parameter: the reference value of $10^{-19}\,\dens$, and three lower values at $3\times10^{-20}\,\dens$, $10^{-20}\,\dens$ and $3\times10^{-21}\,\dens$.

Fig.~\ref{fig: app_B} shows the variation with the persistence threshold of the extracted cumulative distribution function of $\theta_{B\varphi}$. As a reference, we display the snapshots taken at $\SFE=3\%$.
We can see that the parameters do modify the distributions. 
The general features we discussed in this work remain unchanged. \simBB filaments continue to prefer perpendicularity relative to B (this is particularly visible at lower angles), while \simA filaments remain consistently parallel to it. In fact, by construction, the filaments extracted with high persistence thresholds are always a subsample from the set extracted at lower values of the parameter. We note that as the threshold decreases and more filaments are retrieved in the domain, the distribution of $\theta_{B\varphi}$ (Fig.~\ref{fig: app_B}) seems to reach a convergence. However, it is also true that the distributions obtained (especially in \simBB) get closer to the random distribution (solid black line) as we lower persistence thresholds. 

This connects to what was discussed in Sec.~\ref{sec: methods_fil_extraction}. Algorithms based on local derivatives are effective in finding real filamentary structures, but are also sensitive to the noise of the density field. In a turbulent environment such as that of star-forming clouds, random fluctuations are frequent, and only a high persistence threshold can ensure the filtering out of the false positive filaments returned by the code. Fig.~\ref{fig: filamentary_at_thrs} proves that, displaying how the extracted skeleton changes as a function of the parameter for the snapshots at $\SFE=3\%$ in \simBB (top) and \simBB (bottom). Adopting high persistence values ($10^{-19}\,\dens$ and $3\times10^{-20}\,\dens$, first and second column respectively), provides a skeleton which is consistent with the density structure that can be observed `by eye'. Already at $10^{-20}\,\dens$ (third column), we can tell that most of the structures are not real, while at $3\times10^{-21}\,\dens$ (last column) we are entirely dominated by noise. 
Being interested in sampling the filaments that take part in the SF process, we opted for a persistence threshold closer to the critical density for isothermal filaments (see text, also \citealt{Andre2014}), for which we can be confident that the detected filaments are real features.

\section{Filament properties before the onset of star formation}
\label{sec: appendix before SF}
To assess whether the filament properties extracted from the skeleton at the onset of SF (SFE=0.1\,\%) can be linked to those during the actual filament formation phase, we conducted an additional analysis on the epochs prior to the formation of the first star. While this work mainly focuses on the properties the filaments develop during their star-forming phase, it is still instructive to explore whether some of these characteristics are already imprinted earlier to get insights on the filament formation process. 

However, during the assembly phase, the maximum density reached by filaments is relatively low ($\lesssim10^{4-5}\,\ndens$), and the high persistence threshold adopted leads to extremely poor statistics. Therefore, we lowered the threshold when analysing earlier snapshots of the cloud. Nonetheless, selecting an appropriate threshold at each snapshot introduces an element of arbitrariness. Rather than choosing it by eye based on whether the resulting skeleton looked too noisy or sparse, we opted for a more quantitative approach. For each snapshot, we looked for a threshold producing a skeleton with a total length (within $10\%$) comparable with that obtained at $\SFE=0.1\%$.

At the simulation epoch with $\SFE=0.1\%$ (leftmost column of Fig.~\ref{fig: cloud_simulation}), the two skeletons have a total length of $\approx33\,\pc$ in \simBB and $13\,\pc$ in \simA. Among the previous snapshots, we selected the ones at $0.3,\,0.6,\,0.9\,\Myr$ before the formation of the first sink particle ($\tau_\mathrm{sf}$). For these, the selected persistence thresholds were 
$8, \,10$ and $40\%$ of the threshold adopted in the text for the strongly magnetised cloud and $9, \,30$ and $50\%$ for the weakly magnetised run (from the earliest snapshot to the oldest).

\begin{figure*}[h!]
    \centering
    \includegraphics[width=0.9\linewidth]{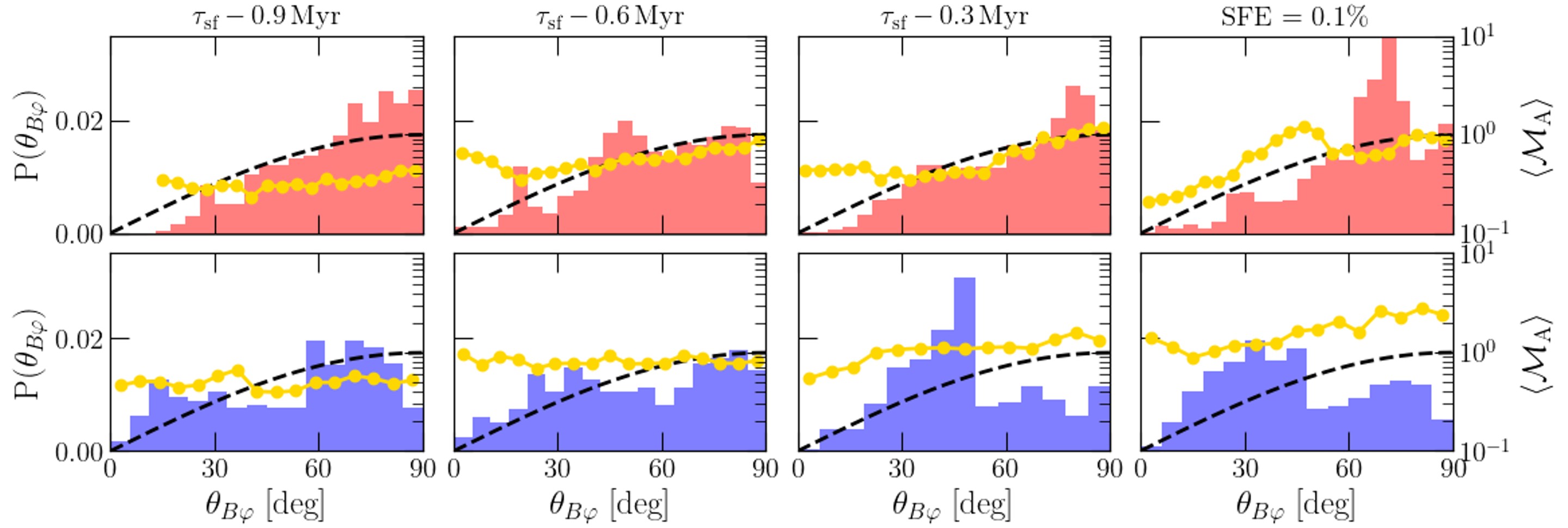}
    \caption{
    Distributions of the relative angle between the magnetic field and the filament ($\theta_{B\varphi}$) in \simBB\ (top) and \simA\ (bottom), at different snapshots prior to the onset of the SF ($\tau_\mathrm{sf}$), corresponding to the $SFE=0.1\%$ stage. The black dashed lines mark the random distribution for the relative orientation of two independent vectors in 3D. The yellow dotted lines mark the average Alfvén Mach number $\mathcal{M}_\mathrm{A}$ for filament portions at a given $\theta_{B\varphi}$ (scale on the right axis).
\label{fig: app thetaBphi}}
\end{figure*}

\begin{figure*}[h!]
    \centering
    \includegraphics[width=0.9\linewidth]{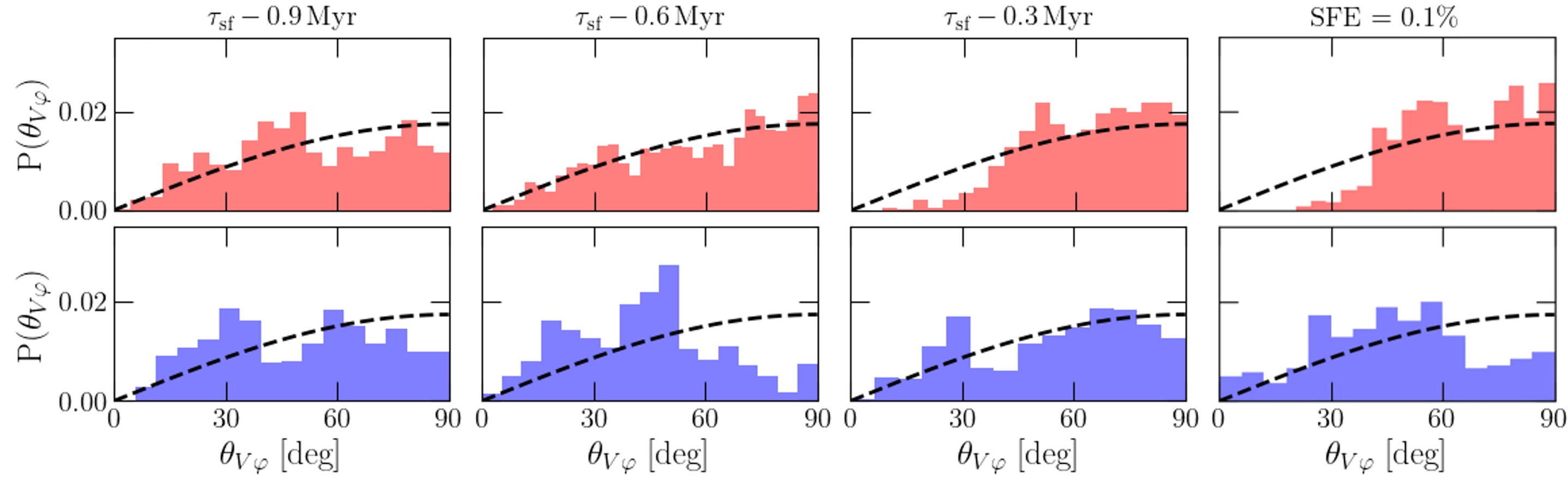}
    \caption{
    Evolution of the distribution of the relative angle between the velocity vector of the filament crest and the filament direction ($\theta_{V\varphi}$) in \simBB\ (top) and \simA\ (bottom), at different snapshots prior the onset of the SF ($\tau_\mathrm{sf}$). The black dashed lines mark the random distribution for the relative orientation of two independent vectors in 3D. 
\label{fig: app thetaVphi}}
\end{figure*}

We chose to perform a comparison with two of the main statistics discussed in the text, the distributions of $\theta_{B\varphi}$ (Fig.~\ref{fig: app thetaBphi}) and $\theta_{V\varphi}$ (Fig.~\ref{fig: app thetaVphi}). Both differences and similarities can be detected. From Fig.~\ref{fig: app thetaBphi}, we see that the orientation trend at $\SFE=0.1\%$ is inherited from earlier stages, supporting the idea that the alignment is imprinted by the formation mechanism and the environmental properties in which these structures form. Moreover, we find a further resemblance in the behaviour of $\machav$, which retains its dependence on $\theta_{B\varphi}$.

Nonetheless, in \simBB these stages display a closer resemblance to the random distribution. Part of this could be explained with the lower persistence threshold, which might be more sensitive to noisy fluctuations, and part with the higher turbulence level -- that has yet to dissipate -- originating filaments from random density fluctuation seeds. A visual inspection of the skeletons (not shown here) favours the second option, but a more accurate analysis is required to assess whether this initial similarity in $\mu=2$ and $\mu=8$ distributions has a physical origin. However, as stated in the text, a detailed analysis of these stages is beyond the purpose of the current study, which focuses on the evolution of filament properties.

A more significant difference can be detected in the $\theta_\mathrm{V\varphi}$ distribution. In \simA (bottom) the distribution roughly behaves as in the one seen at $\SFE=0.1\%$, while in \simBB (top) we see a more gradual progression towards perpendicular alignments. 
In summary, we find that the filament properties at earlier stages quite resemble those at the onset of SF, allowing us to infer to some extent, some conclusion on the dynamical formation process of filaments based on the properties at the snapshot $\SFE=0.1\%$ studied in the text.

\end{appendix}

\end{document}